\documentclass[a4paper,fleqn,usenatbib]{mnras}

\usepackage{mathptmx}

\usepackage[T1]{fontenc}
\usepackage{ae,aecompl}

\usepackage{graphicx}	
\usepackage{amsmath}	
\usepackage{amssymb}	
\usepackage{multicol}        
\usepackage{bm}		
\usepackage{pdflscape}	
\usepackage{upgreek}

\newcommand{\kms}{\,km\,s$^{-1}$} 
\newcommand{\kkms}{K\,km\,s$^{-1}$} 
\newcommand{\co}{$^{12}$CO}
\newcommand{\coa}{$^{13}$CO ($J=3\rightarrow 2$)}
\newcommand{\cob}{C$^{18}$O ($J=3\rightarrow 2$)}
\newcommand{\cohrs}{$^{12}$CO ($J=3\rightarrow 2$)}
\newcommand{\cogrs}{$^{13}$CO ($J=1\rightarrow 0$)}

\newcommand{\tast}{$T_\mathrm{A}^*$}
\newcommand{\tmb}{$T_\mathrm{mb}$}

\title[CHIMPS]{CHIMPS: the $^{13}$CO/C$^{18}$O ($J=3\rightarrow 2$) Heterodyne Inner Milky Way Plane Survey}

\author[A. J. Rigby et al.]{A. J. Rigby$^1$\thanks{E-mail: ajrigby24@gmail.com}, T. J. T. Moore$^1$, R. Plume$^2$, D. J. Eden$^1$, J. S. Urquhart$^3$,
\newauthor M. A. Thompson$^4$, J. C. Mottram$^5$, C. M. Brunt$^6$, H. M. Butner$^{7}$,
\newauthor J. T. Dempsey$^{8}$, S. J. Gibson$^{9}$, J. Hatchell$^{6}$, T. Jenness$^{8,10}$, N. Kuno$^{11}$,
\newauthor S. N. Longmore$^1$, L. K. Morgan$^{12}$, D. Polychroni$^{13}$, H. Thomas$^{8}$, G. J. White$^{14,15}$,
\newauthor M. Zhu$^{16}$\\
$^{1}$Astrophysics Research Institute, Liverpool John Moores University, IC2, Liverpool Science Park, 146 Brownlow Hill, Liverpool L3 \\ 5RF, United Kingdom\\
$^{2}$Department of Physics and Astronomy, University of Calgary, 2500 University Drive NW, Calgary, AB T2N 1N4, Canada \\
$^{3}$Max-Planck-Institut f\"ur Radioastronomie, Auf dem H\"ugel 69, D-53121 Bonn, Germany\\
$^{4}$Centre for Astrophysics Research, Science \& Technology Research Institute, University of Hertfordshire, College Lane, Hatfield, \\ Herts AL10 9AB, UK \\
$^{5}$Leiden Observatory, Leiden University, PO Box 9513, 2300 RA Leiden, The Netherlands\\
$^{6}$School of Physics, University of Exeter, Stocker Road, Exeter, EX4 4QL, UK\\
$^{7}$Department of Physics and Astronomy, James Madison University, MSC 4502, 901 Carrier Drive, Harrisonburg, VA 22801, USA\\
$^{8}$Joint Astronomy Centre, 660 N. A'ohoku Place, University Park, Hilo, HI 96720, USA\\
$^{9}$Department of Physics and Astronomy, Western Kentucky University, 1906 College Heights Blvd., Bowling Green, KY 42101, USA\\
$^{10}$ Large Synoptic Survey Telescope Project Office, 933 N. Cherry Ave, Tucson, AZ 85721, USA \\
$^{11}$ Division of Physics, Faculty of Pure and Applied Science, Center for Integrated Research in Fundamental Science and Engineering \\(CiRfSE), University of Tsukuba, 1-1-1 Ten-nodai, Tsukuba, Ibaraki, 305-8571, Japan\\
$^{12}$Met Office, FitzRoy Road, Exeter, Devon, EX1 3PB, UK\\
$^{13}$Department of Astrophysics, Astronomy and Mechanics, Faculty of Physics, University of Athens, Panepistimiopolis, 15784 \\Zografos, Athens, Greece\\
$^{14}$RALSpace, Rutherford Appleton Laboratory, Chilton, Didcot, Oxfordshire, OX11 0QX, UK\\
$^{15}$Department of Physics and Astronomy, The Open University, Walton Hall, Milton Keynes, MK7 6AA, UK\\
$^{16}$National Astronomical Observatories, Chinese Academy of Science, 20A Datun Road, Chaoyang District, Beijing 100012, China\\}

\date{Accepted 2015 November 27. Received 2015 November 03; in original form 2015 September 16.}

\pubyear{2015}

\begin{document}
\label{firstpage}
\pagerange{\pageref{firstpage}--\pageref{lastpage}}
\maketitle

\begin{abstract}
We present the  $^{13}$CO/C$^{18}$O ($J=3\rightarrow 2$) Heterodyne Inner Milky Way Plane Survey (CHIMPS) which has been carried out using the Heterodyne Array Receiver Program on the 15\,m James Clerk Maxwell Telescope (JCMT) in Hawaii. The high-resolution spectral survey currently covers $|b| \leq 0\fdg 5$ and $28\degr \lesssim l \lesssim 46\degr$, with an angular resolution of 15 arcsec in 0.5\kms\ velocity channels. The spectra have a median rms of $\sim$ 0.6\,K at this resolution, and for optically thin gas at an excitation temperature of 10 K, this sensitivity corresponds to column densities of $N_{\mathrm{H}_{2}} \sim 3 \times 10^{20}\,$cm$^{-2}$ and $N_{\mathrm{H}_{2}} \sim 4 \times 10^{21}\,$cm$^{-2}$ for $^{13}$CO and C$^{18}$O, respectively. The molecular gas that CHIMPS traces is at higher column densities and is also more optically thin than in other publicly available CO surveys due to its rarer isotopologues, and thus more representative of the three-dimensional structure of the clouds. The critical density of the $J=3 \rightarrow 2$ transition of CO is $\gtrsim 10^{4}$\,cm$^{-3}$ at temperatures of $\leq 20$\,K, and so the higher density gas associated with star formation is well traced. These data  complement other existing Galactic plane surveys, especially the JCMT Galactic Plane Survey which has similar spatial resolution and column density sensitivity, and the \textit{Herschel} infrared Galactic Plane Survey. In this paper, we discuss the observations, data reduction and characteristics of the survey, presenting integrated emission maps for the region covered. Position velocity diagrams allow comparison with Galactic structure models of the Milky Way, and while we find good agreement with a particular four arm model, there are some significant deviations.
\end{abstract}



\section{Introduction}

Molecular clouds make up the coldest and densest regions of the interstellar medium (ISM) and are the sites of star formation. Their composition is hierarchical \citep[e.g.][]{Rosolowsky+08} and possibly fractal or multifractal \citep[e.g.][]{Stutzki+98}, with clouds containing a variety of increasingly dense sub-structures on smaller spatial scales. These structures have densities ranging from a few hundred cm$^{-3}$ in molecular clouds up to $\sim 10^{6}$ cm$^{-3}$ in the dense cores which are the birth places of stars. The earliest stages of star formation can be scrutinized using observations of these increasingly dense molecular components which trace the gravitational collapse of the molecular cloud.

Although molecular clouds consist chiefly of molecular hydrogen and inert atomic helium ($\sim\,$27\% by mass; \citealt{Kauffmann+08}), at typical molecular cloud temperatures ($T_{K} \sim 10$ K)  these species are practically invisible; H$_{2}$ molecules do not possess a permanent dipole moment, and so do not radiate via the electric dipole rotational transitions which are easily excited in other ISM molecules. In addition, the lowest lying quadrupole transitions of H$_{2}$ have small transition probabilities and require excitation temperatures much higher than those typically found in the cold ISM. Carbon monoxide (CO) is the second most abundant molecule in the ISM, being almost ubiquitous with H$_{2}$, and has a fundamental rotational transition with an energy of $E/k \approx 5.5 $\,K \citep{Bolatto+13}. CO is, therefore, an ideal tracer for molecular gas, and has a number of rotational transitions which can be observed using submillimetre telescopes. 

The most common CO isotopologue is $^{12}$C$^{16}$O (hereafter `\co'), and its submillimetre $J=3 \rightarrow 2$ rotational emission line has a critical density of $1.6 \times 10^{4}$ cm$^{-3}$ at 10\,K (using recent figures from LAMDA; \citealt{Schoier+05}), meaning that this transition is sensitive to higher density gas than preceding surveys in $J=1 \rightarrow 0$ such as the Boston University and Five College Radio Astronomy Observatory Galactic Ring Survey \citep[GRS;][]{Jackson+06}. Observations of the $J=3 \rightarrow 2$ transition in different isotopologues of CO such as $^{13}$CO and C$^{18}$O, which have lower fractional abundances than \co\ are able to trace the gas in these densest structures to high optical depths; $^{13}$CO is $\sim 50$--$100$ times less abundant than \co\ \citep{Schoier+02} and C$^{18}$O is roughly 10 times less abundant than $^{13}$CO \citep{Hogerheijde+98}. Observations of a particular molecular transition in multiple isotopologues (e.g. $^{13}$CO and C$^{18}$O) allow optical depths to be derived, and when combined with observations of the same species in multiple transitions, excitation temperatures may also be directly determined \citep[e.g.][]{Polychroni+12}. Column densities can be determined from the optical depths and excitation temperatures, and if distances can be determined, then these parameters are also sufficient to allow masses and mean densities to be derived.

High--resolution observations of optically thin dense gas emission lines may allow kinematic velocities to be assigned to star-forming regions seen in continuum data with less ambiguity than other tracers. Observations in \cohrs, for example, often display multiple complex emission features along lines of sight within the Galactic plane, corresponding to emission from separate structures at different distances along the line of sight. \citet[see fig. 1]{Eden+12} demonstrate that spectra from also \cogrs\ suffer from similar problems, with additional confusion from more diffuse gas. Spectra of \coa\ generally display only a single narrow emission feature, and hence are useful in estimating which spectral features are likely to correspond to star-forming sites. Kinematic velocities can enable source distances to be calculated by assuming a Galactic rotation curve \citep[e.g.][]{Brand+Blitz93,Russeil03}. 

Position--velocity diagrams from molecular gas surveys are an excellent diagnostic to use to test and refine models of Galactic structure \citep[e.g.][]{Dame+01}. While the debate as to the number of spiral arms the Milky Way appears to have settled on four \citep[e.g.][]{Hou+Han14,Urquhart+14} main spiral arms, the precise positions of these arms need further refining and there are still a large number of other features such as spurs to be explained and incorporated into models. There is also evidence that while molecular gas, star formation and young stars trace the four spiral arms, older stars such as K and M giants are better described in terms of a two-arm model \citep[][and references therein]{Steiman-Cameron+10}. A recent study by \citet{Ragan+14} identified giant molecular filaments in the Galactic plane, and by using position--velocity diagrams postulated that these structures are largely inter-arm in nature, and may be the analogues of spurs seen in nearby spiral galaxies \citep[e.g.][]{Elmegreen+80}. A high contrast gas tracer measured over a significant section of the Galactic plane will allow details of models of spiral discs, and especially those with synthetic position--velocity observations \citep[e.g.][]{Duarte-Cabral+15,Pettitt+15}, to be thoroughly tested and constrained.

In this paper, we present data from an 18 square-degree region of  the $^{13}$CO/C$^{18}$O ($J=3 \rightarrow 2$) Heterodyne Inner Milky Way Plane Survey (CHIMPS), which is now publicly available. CHIMPS serves to complement a number of Galactic plane surveys in various different CO isotopologues and transitions that have become publicly available in recent years; the aforementioned GRS \citep{Jackson+06} mapped the region $18\degr \leq l \leq 55\fdg 7$ and $|b| \leq 1\degr$ in $^{13}$CO ($J=1 \rightarrow 0$) with an angular resolution of $46$ arcsec; the \cohrs\ High-Resolution Survey of the Galactic plane (COHRS; \citealt{Dempsey+13}) is ongoing and has currently charted an area of $17\fdg 5 \leq l \leq 50 \fdg 25$ with a width of $|b| \leq 0 \fdg 25$, and with $|b| \leq 0\fdg5$ for two small segments with an angular resolution of $14$ arcsec. CHIMPS also serves to complement the growing number of Galactic plane continuum surveys at sub/millimetre and infra-red wavelengths such as the JCMT Galactic Plane Survey \citep[JPS;][]{Moore+15}, the Bolocam Galactic Plane Survey (BGPS; \citealt{Aguirre+11}), the APEX Telescope Large Area Survey of the Galaxy (ATLASGAL; \citealt{Schuller+09}), the \textit{Herschel} infrared Galactic Plane Survey (Hi-GAL; \citealt{Molinari+10a}), \textit{ Wide-field Infrared Survey Explorer} (\textit{WISE}; \citealt{Wright+10}), \textit{Spitzer}'s Galactic Legacy Infrared Mid-Plane Survey Extraordinaire (GLIMPSE; \citealt{Benjamin+03,Churchwell+09}) and MIPSGAL \citep{Carey+09}. In addition to the molecular gas, infra-red and sub/millimetre surveys, the Coordinated Radio and Infrared Survey for High-Mass Star Formation \citep[CORNISH;][]{Hoare+12,Purcell+13} at $5\,$GHz has catalogued ultra-compact \ion{H}{ii} regions, which are indicators of massive star formation, over a congruent area.

In Section \ref{sec:obs} we describe the CHIMPS observations and the data reduction process. Section \ref{sec:dataoverview} provides an overview of the data and discusses their properties. Details of how to access the CHIMPS data are presented in Section \ref{sec:access}. We compare CHIMPS with other available molecular gas surveys in Section \ref{sec:comparison}, and present close-ups of some example CHIMPS fields in Section \ref{sec:previews}, summarizing in Section \ref{sec:summary}.

\section[]{Observations and data reduction}\label{sec:obs}

\subsection{Observations}

CHIMPS is a spectral survey covering the $J=3\rightarrow 2$ rotational transitions of  $^{13}$CO at $330.587\,$GHz and C$^{18}$O at $329.331\,$GHz. The observations were made using the Heterodyne Array Receiver Program (HARP; \citealt{Buckle+09}) on the $15 \,$m James Clerk Maxwell Telescope (JCMT) in Hawaii. The observations cover approximately 18 square degrees in the region $27 \fdg 5 \lesssim l \lesssim 46 \fdg 4$ and $|b| \leq 0 \fdg 5$, and were taken over a total of eight semesters at JCMT, beginning in 2010 March. The most recent data presented here were taken in 2014 June, and the proposal IDs are: m10ac06, m10au13, m10bu28, m11au05, m12bc19, m12bu37, m13au31, m13bu28, s13bu03 and s14au04.

HARP is a 16-receptor focal-plane array receiver operating over a submillimetre frequency range of $325 - 375\,$GHz. The receptors are superconductor--insulator--superconductor heterodyne detectors arranged in a $4 \times 4$ grid, each separated by $30$ arcsec on the sky. The Auto-Correlation Spectral Imaging System \citep{Buckle+09} backend was used in conjunction with HARP and, configured to use a 250 MHz bandwidth with 4096 frequency channels of width $61.0\,$kHz. The velocity width per channel is 0.055\kms\, giving each CHIMPS observation $\sim 200$\kms of usable velocity coverage. In the kinematic local standard of rest (LSRK) the velocity window was placed at $-50$ to 150\kms\ at $l = 28\degr$, and shifts with increasing Galactic longitude to $-75$ to 125\kms\ at $l = 46\degr$ in order to follow the Galactic velocity gradient. This range covers expected velocities of the regions associated with the Scutum-Centaurus tangent, and the Sagittarius, Perseus and Norma arms.

The observations were taken in a position-switching raster (on-the-fly) mode with off-positions measured below the Galactic plane with a latitude offset of $\Delta b= -1 \fdg 5$ for each observation. This observation mode scans across the area of sky by the desired width filling the image with the first few rows of pixels. When the scan reaches the edge of the sky region, the array is shifted in a direction perpendicular to the scan direction before scanning over the field again in the reverse direction. In this way, each point of sky is covered by multiple receptors. This process is repeated until the required area of sky is covered, and a second scan is then made by passing over the same area with a scan direction orthogonal to that of the first scan. A 1/2 array scan spacing was used, which shifts the array by half of its width in a direction perpendicular to the scan direction when it completes each row, before the scan direction is reversed. The raw data are written continuously as the telescope scans, in a time series format. This results in a sample spacing of $7.3$ arcsec which, in conjunction with a $0.25\,$s sample time, produces data cubes covering an area of $\sim$ 21 arcmin $ \times$ 21 arcmin in approximately one hour. A small number of observations, however, are slightly larger or smaller in size as discussed later on in this section.  

As part of the standard operating procedure at JCMT, pointing accuracy is checked between most observations, and is generally found to be approximately 2 arcseconds in both azimuth and elevation. Tracking accuracy is better than 1 arcsecond over the course of a typical $\sim 1$ hour observation. The spectra are calibrated as the observations are made, using the three-load chopper-wheel method of \citet{Kutner+Ulich81}. Intensities are thereby placed on the \tast\ (corrected antenna temperature) scale, which corrects for atmospheric attenuation, ohmic losses within the telescope, and rearward scattering and spillover. This \tast\ scale is then calibrated absolutely by observations of spectral standards (listed online\footnote{http://www.eaobservatory.org/jcmt/instrumentation/heterody\\ne/calibration/}) that are carried out on a nightly basis. Calibrated peak and integrated intensities of the standards must fall within 20\% of the standard values, or else the receiver is re-tuned and calibration is repeated. The \tast\ intensities can be converted to main beam brightness temperature (\tmb) by using the relation \tmb\ $=$ \tast/$\eta_\mathrm{mb}$ adopting the mean detector efficiency $\eta_\mathrm{mb}=0.72$ \citep{Buckle+09}. All intensities reported in this paper are on the \tast\ scale unless stated otherwise.

The tiling pattern for the observations varies over three sections. In the section spanning $27 \fdg 5 \lesssim l \lesssim 32 \fdg 8$, the cubes were observed such that the edges of the map are aligned in the equatorial coordinate system. For longitudes of $32 \fdg 8 \lesssim l \lesssim 44 \fdg 1$, the cubes have the same dimensions as the lower longitude section, but are parallel to Galactic longitude and latitude. This tiling pattern was more efficient since no time was spent observing latitudes $|b| > 0 \fdg 5$. The change in tiling pattern was due to an update to the observation setup for HARP raster maps which made it possible to observe square maps aligned with Galactic coordinates. The final $44 \fdg 1 \lesssim l \lesssim 46 \fdg 4$ section was observed contemporaneously with the lowest longitude section, and consequently the observation edges are aligned with the equatorial gridlines. In the latter section the cubes also have slightly different dimensions; 18 of the cubes here measure approximately 22 arcmin along each side, and 10 cubes measure $\sim$ 7.5 arcmin along each side; the smaller observations were to fill holes which were not covered by the original tiling pattern.

\subsection{Data reduction}

The raw time series data were reduced using the {\sc orac-dr} data reduction pipeline \citep{Jenness+15} which is built on the Starlink \citep{Currie+14} packages {\sc cupid} \citep{Berry+07}, {\sc kappa} \citep{Currie+08} and {\sc smurf} \citep{Jenness+08}; specifically, the {\sc narrowline} reduction recipe was used, which is optimized for Galactic targets with narrow line widths (compared to the bandwidth) and small velocity gradients. The reduction pipeline transforms the raw time series spectra into spectral data cubes with longitude, latitude and velocity $(l, b, v)$ axes. We refer the reader to \citet{Dempsey+13} and \citet{Jenness+15} for more detailed descriptions of the pipeline. The default quality assurance parameters were used as listed in table 2 of \citet{Dempsey+13}. The pixel size used is 7.6 arcsec, half of the beamwidth at this frequency and, to increase signal-to-noise, the spectral axis was re-binned into 0.5\kms\ velocity channels. Baseline subtractions were carried out using a fourth-order polynomial fit which was found to have sufficient flexibility to fit both linear and typical non-linear baselines well. Such bad baselines may result from external interference, for example \citep[cf.][]{Currie13}. Prior to reduction, an average spectrum was generated for each time series observation by integrating over the time and position axes to determine the velocity region for any strong emission. These velocity regions were then masked out for the baseline subtraction by the software in order to avoid fitting the baseline polynomial to any broad emission features. The {\sc orac-dr} parameters are listed in Appendix \ref{app:orac-pars}.

The reduced data cubes each contain a variance array component determined for each spectrum from the system noise temperature by the {\sc smurf} utility {\sc makecube} within the reduction pipeline. Upon output from {\sc orac-dr}, the reduced cubes have undersampled edges caused by the change in direction of the scanning pattern when generating the raster maps, which also have low-signal to noise. The cubes are cropped to remove these unwanted edge features. After cropping, there is a small overlap region (typically $\approx$ 1 arcmin) between adjacent tiles that results in a reduced noise level when adjacent tiles are mosaicked (see Fig. \ref{fig:noisemap}).

There are a number of cases where the observation in a particular location has been repeated, and the duplicate observations were co-added using the {\sc mosaic\_jcmt\_images} recipe from Starlink's {\sc picard} package \citep{Gibb+13}, which is contained within {\sc orac-dr}. We have made available all files that make up these combined cubes, should the user wish to co-add them in a different way, or use a single observation.

Additionally, a number of data cubes were taken when a several of the 16 HARP receptors were unusable; sometimes with as few as 11 active receptors. If any further receptors are rejected by {\sc orac-dr}, the reduced data cubes may contain locations with no valid spectra. This effect results in data cubes containing a regular grid of blank spectra at the particular locations which received no sampling. These blank voxels (three-dimensional pixels) were filled in using an interpolation routine ({\sc kappa:fillbad}) which estimates a voxel value from adjacent voxels in the $l$--$b$ plane. These interpolated spectra tend to have high variance values.

Throughout this paper, we refer to three-dimensional $(l,b,v)$ pixels as `voxels', and the term `pixels' is used to describe array elements making up either a two-dimensional $l$--$b$ image, or as the elements of an $l$--$b$ plane from an $(l,b,v)$ cube.

\section{The data}\label{sec:dataoverview}

\subsection{Overview}
\begin{figure}
\centering
\includegraphics[width=0.47\textwidth]{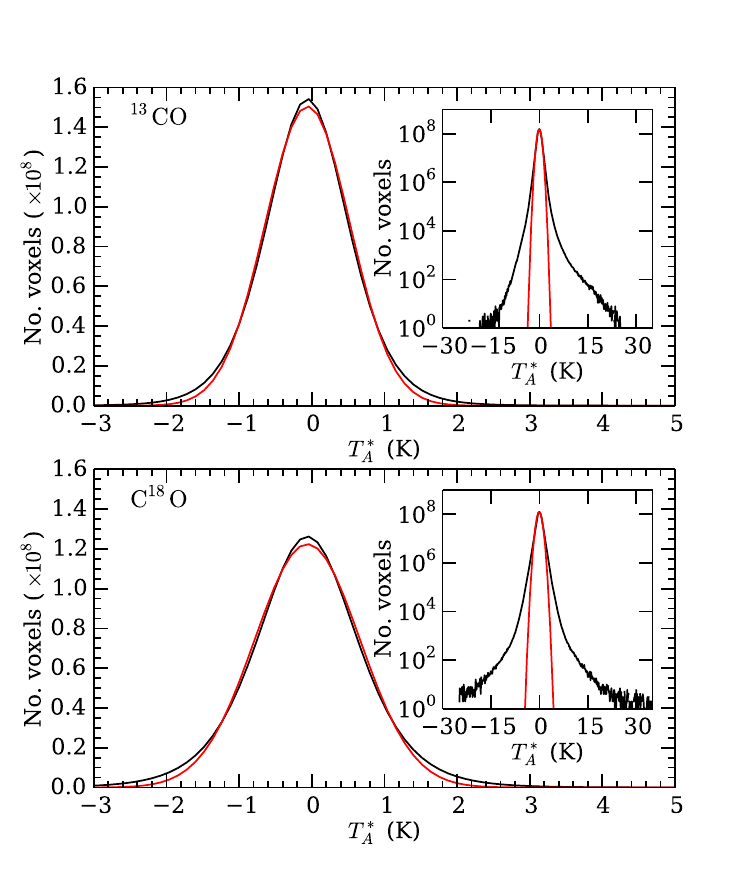}
\caption{Histogram of all voxels in CHIMPS for $^{13}$CO (top) and C$^{18}$O (bottom). The red lines show the Gaussian fits with the functions $1.51 \times 10^{8} \exp[-($\tast$ + 0.06)^{2}/2 \times 0.58^{2}]$ and $1.22 \times 10^{8} \exp[-($\tast$+ 0.06)^{2}/2 \times 0.73^{2}]$ for $^{13}$CO and C$^{18}$O, respectively. The bin width is 0.12$\,$K. The insets show the Gaussian fits on a logarithmic scale.}\label{fig:allvoxels}
\end{figure}

The CHIMPS survey data presented in this paper cover a total of approximately 18 square degrees. A histogram of all voxel values in both isotopologues is shown in Fig. \ref{fig:allvoxels}. The voxel values can be modelled as being normally distributed about  a mean value of $-$0.06\,K in both cases, with a standard deviation of 0.6 and $0.7\,$K in the $^{13}$CO and C$^{18}$O data, respectively. For optically thin gas at an excitation temperature of 10$\,$K (typical of molecular clouds; e.g. \citealt{Polychroni+12}), these sensitivities correspond to gas column densities of $N_{^{13}\mathrm{CO}} \sim 3 \times 10^{14}\,$cm$^{-2}$ and $N_{\mathrm{C}^{18}\mathrm{O}} \sim 4 \times 10^{14}\,$cm$^{-2}$, or  $N_{\mathrm{H}_2} \sim 3 \times 10^{20}\,$cm$^{-2}$ and $N_{\mathrm{H}_2} \sim 4 \times 10^{21}\,$cm$^{-2}$ for \coa\ and \cob, respecitvely, assuming abundance ratios of $^{12}\mathrm{CO}/^{13}\mathrm{CO}=77$ \citep{Wilson+Rood94}, $^{12}\mathrm{CO}/\mathrm{H}_2 \sim 8.5 \times 10^{-5}$ and $\mathrm{C}^{18}\mathrm{O}/\mathrm{H}_2 \sim 1.7 \times 10^{-7}$ \citep{Frerking+82}. For comparison, a higher excitation temperature of $30\,$K would imply a sensitivity to corresponding to a column densities of $N_{^{13}\mathrm{CO}} \sim 1 \times 10^{14}\,$cm$^{-2}$.

There is a strong wing towards the higher positive brightness temperatures in the $^{13}$CO distribution which can be identified as voxels containing emission, and a smaller wing extends out to negative antenna temperatures. The former is much stronger in $^{13}$CO than C$^{18}$O where emission is weaker. The negative wings can be attributed to those observations which have significantly higher-than-average noise levels. The overall distribution is the convolution of the noise distributions for each individual observation, with the addition of detected emission in the positive antenna temperature wing. The 330$\,$GHz band lies on the edge of an atmospheric absorption feature \citep[fig. 20]{Buckle+09}, whereby transmission is lower at lower frequencies; as the lower frequency emission line, the C$^{18}$O data suffer more from the resulting attenuation and hence have broader noise wings in its voxel distribution.

\begin{figure}
\centering
\includegraphics[width=0.47\textwidth]{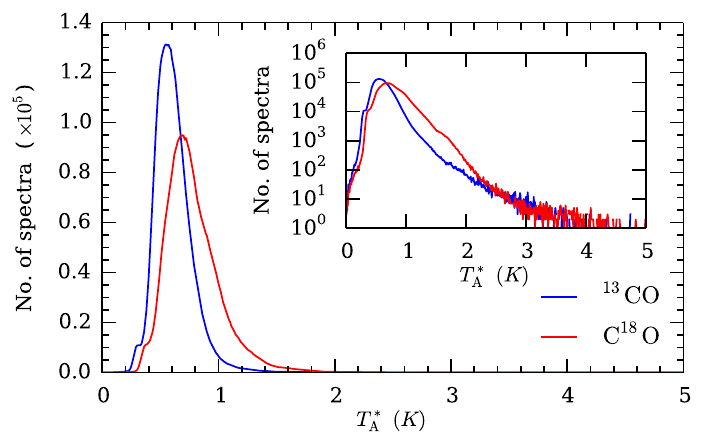}
\caption{Histograms of the noise values in the CHIMPS data. The blue line shows the noise values for the \coa\ data while the red line shows the noise values for the \cob\ data. The bin width is 0.01$\,$K. The inset shows the same distributions on a logarithmic scale.} \label{fig:noisehist}
\end{figure}

\begin{figure*}
\centering
\includegraphics[width=\textwidth]{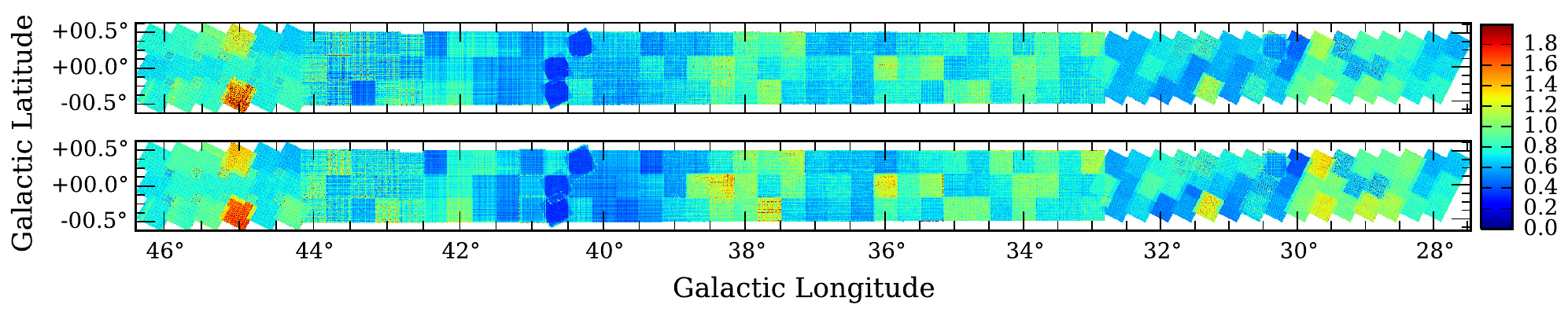}
\caption{Noise (rms) maps for the CHIMPS data. Top: \coa. Bottom: \cob. The intensity scale is in \tast\ (K).} \label{fig:noisemap}
\end{figure*}

A histogram of the rms values of every spectrum in the survey is shown in Fig. \ref{fig:noisehist}. These values were determined by taking the square root of each pixel in the two-dimensional variance arrays that are produced for each observation in the data reduction process. Both distributions peak at values close to the standard deviations of the normal distributions in Fig. \ref{fig:allvoxels}. The rms noise map for each CHIMPS isotopologue is shown in Fig. \ref{fig:noisemap}. The variation of noise across the map is caused by a combination of varying weather conditions, airmasses and variations in the numbers of active receivers on the HARP instrument over the course of the observations. It is also possible to see the lower noise where duplicated or repeated observations have been co-added.

\subsection{Extracting the emission}\label{sec:extraction}

The fully reduced $(l, b, v)$ data cubes contain a significant number of emission-free voxels since the bandwidth is much greater than the velocity width of emission features, even in the brightest regions of the Galactic plane such as the Scutum tangent. In order to avoid integrating large numbers of noise voxels in each spectrum to form an integrated intensity map with a significant noise component, a source extraction was carried out.

To do this, the entire 18 square degrees of CHIMPS were first mosaicked in several sections using {\sc kappa:wcsmosaic} which uses a Lanczos kernel of the form sinc($\uppi x$) sinc ($\uppi k x$), where $x$ is the pixel offset from the input pixel, to assign pixel values in the mosaicked image's pixel grid. A spatial smoothing was then applied using a Gaussian kernel with a full width at half maximum (FWHM) of 3 pixels in order to account for the beam profile as well as a small smoothing effect caused by the re-gridding of pixels in the mosaicking routine, resulting in an effective resolution of 27.4 arcsec. 

A signal-to-noise ratio (SNR) cube of each survey section in both isotopologues was produced using {\sc kappa:makesnr}, which divides the intensity of each voxel by the square root of the variance value of the spectrum to which the voxel belongs. The emission generally occupies a small part of the spectrum, so the fact that the emission is not masked out before calculating the variance is of little consequence. A spatial filtering routine ({\sc cupid:findback}) was next applied to subtract an estimate of the background from each spectrum, and to minimize the regular noise features which appear in the CHIMPS cubes due to variations in sensitivity between receptors which are discussed in Section \ref{sec:dataoverview}.

The source extraction algorithm `FellWalker' \citep{Berry15} in the {\sc cupid} routine {\sc findclumps} was applied to the background-subtracted SNR cubes. For each voxel, FellWalker examines its neighbouring voxels for any higher values, moving to the highest value within the search volume if possible. If no adjacent voxels have a higher value, then the search radius is increased (up to a user-defined maximum search radius), and a jump is made to the new highest voxel value found. When a peak is reached and there are no higher values in the neighbourhood, a clump is defined, and all voxels which lead to that peak are designated as being part of the clump. There is an additional criterion for the minimum number of voxels required for a clump to be defined, in an attempt to reduce false positives from noise spikes, which was set to the minimum allowed value of 16 (corresponding to a cubic source of width 2.5 pixels). FellWalker was chosen for this study over the ClumpFind algorithm  \citep{Williams+94} because comparisons by \citet{Berry15} on a sample of simulated Gaussian clumps found that the FellWalker results are less dependent on the specific parameter settings than for ClumpFind.

Source extraction was carried out on the SNR cubes instead of the intensity cubes so that the effects of the varying background over the 178 individual cubes would not cause either faint sources that have good signal-to-noise in regions of low background to be missed, or false positives to present a significant issue. The background in the original cubes varied significantly between individual observations taken over the course of 4 years due to a varying number of active receptors and the variable weather conditions the data were taken under. A similar approach is used in \citet{Moore+15} who also found that the best results were achieved using FellWalker on SNR maps.

The parameters used for the FellWalker source extraction are listed in Appendix \ref{app:fwpars}. For the extraction of $^{13}$CO sources, the noise level was regarded as all voxels with SNR $< 3$, and sources were required to have a peak with SNR $>5$. Due to comparative rarity of C$^{18}$O compared to $^{13}$CO, the criteria for extraction of C$^{18}$O sources had to be less exacting; the SNR threshold below which voxels are considered noise was lowered 2, though sources were still required to have a peak with SNR $>5$.

Finally, the mask that was produced by FellWalker was applied to the reduced data, effectively removing all voxels that were not identified as emission, resulting in cleaner integrated-emission maps.

\subsection{Integrated position--position maps} \label{sec:emissionlb} \label{sec:emissionmaps}%

\begin{figure*}
\centering
\includegraphics[width=\textwidth]{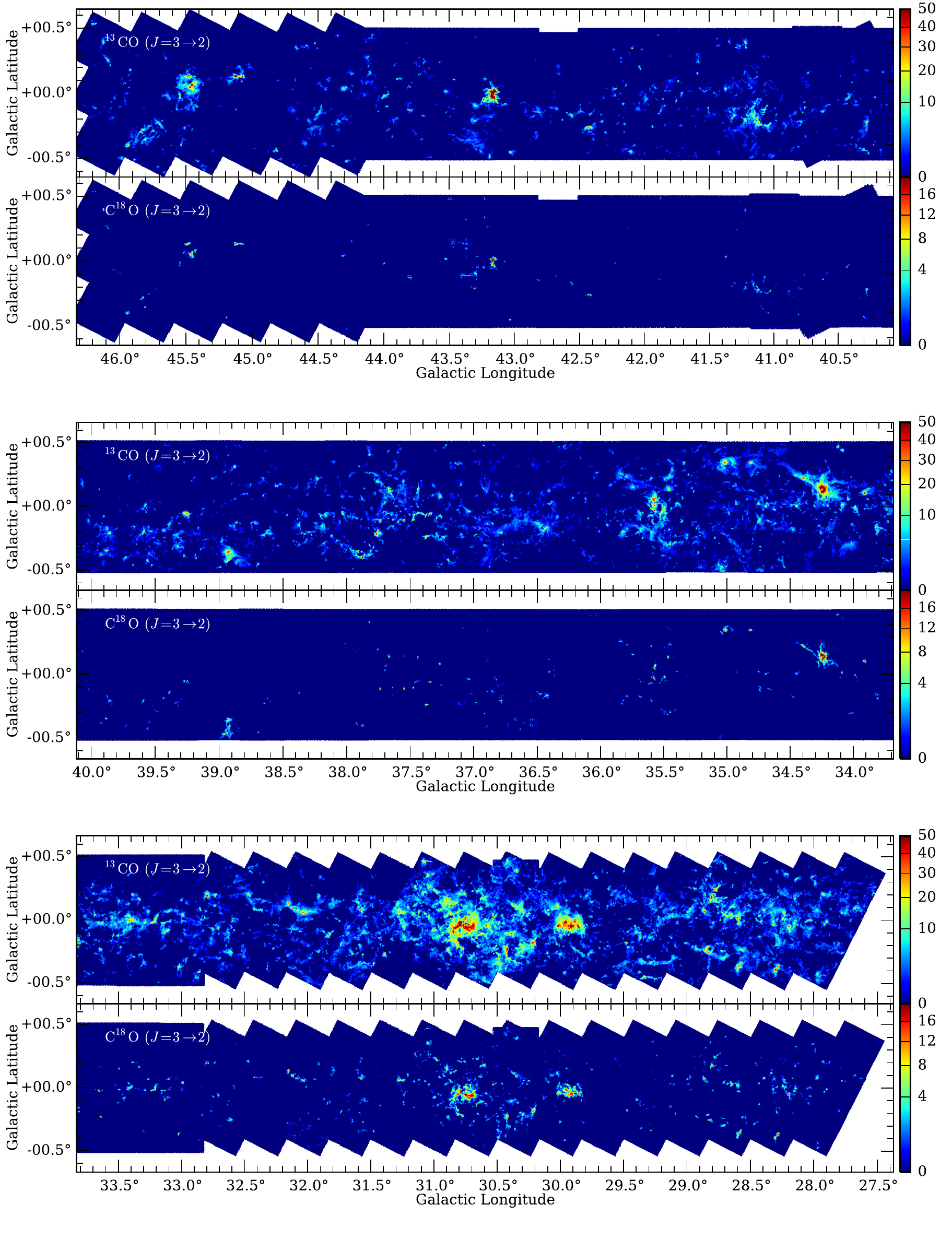}
\caption{The integrated emission (\tast) in CHIMPS. All voxels with an SNR above 5 are included, and any voxels containing emission above an SNR of 3 (in the $^{13}$CO) or 2 (in the C$^{18}$O) which is assigned to a clump with a peak SNR of more than 5 are also included. Each spectrum was integrated over all velocity channels. The units on the intensity scale are K \kms.} \label{fig:integlb}
\end{figure*}

\begin{landscape}
\begin{figure}
\centering
\vspace{2cm}
\includegraphics[width=1.35\textwidth]{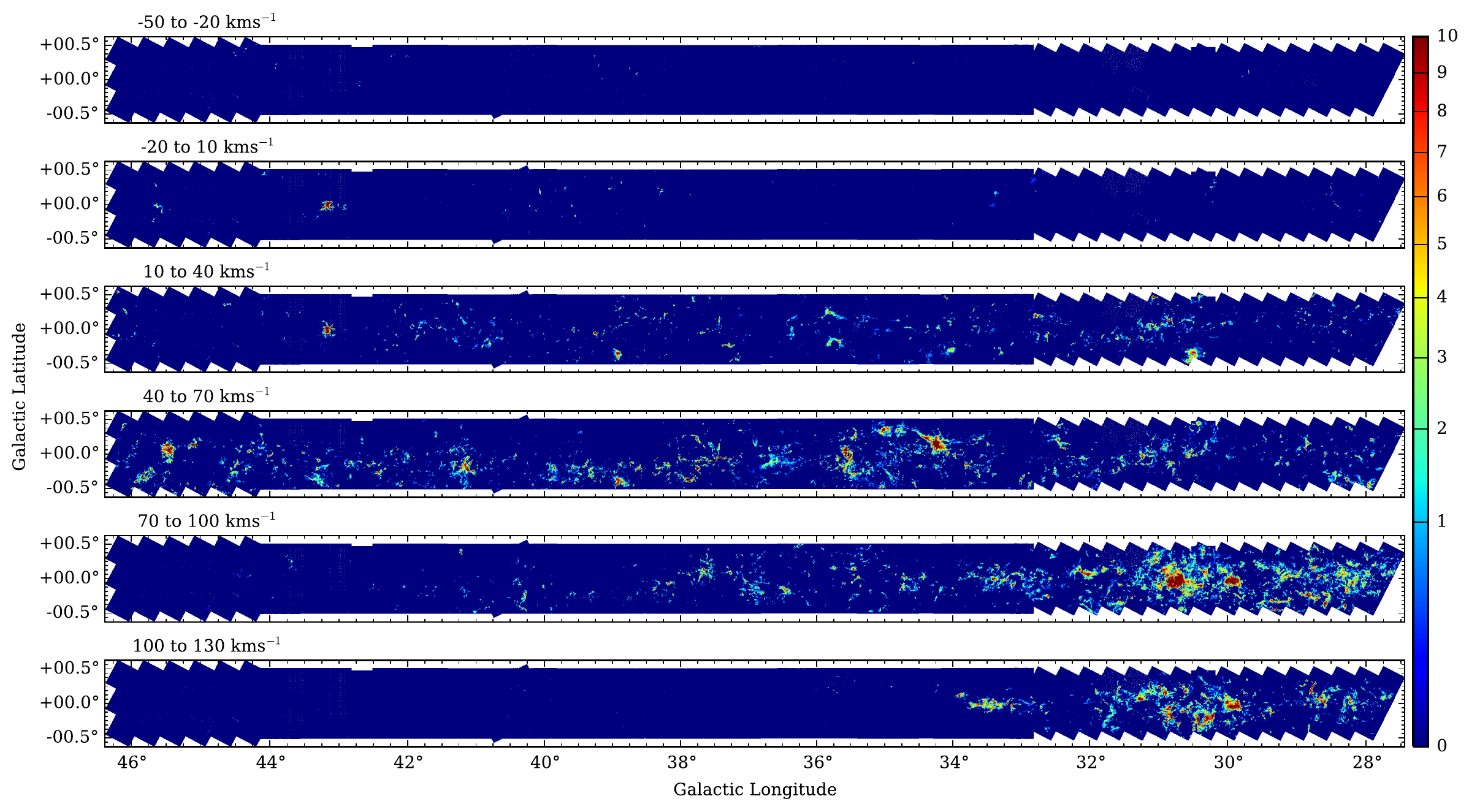}
\caption{Integrated emission (\tast) of the $^{13}$CO data of CHIMPS split into 30\kms\ channels. All emission with an SNR of 5 or higher is shown, and additionally any emission with with an SNR of at least 3, which is related to a clump with a peak SNR of 5 or more. The units of the intensity scale are K\kms.} \label{fig:channelmaps}
\end{figure}
\end{landscape}

Fig. \ref{fig:integlb} shows all emission with SNR $>5$ (measured for each individual spectrum) in the survey, integrated over all velocity channels and in both isotopologues, and additionally any emission with SNR $>3$ in the case of $^{13}$CO, or SNR $> 2$ in the case of C$^{18}$O, which was assigned to a clump. As a result of the FellWalker parameters used, any voxels containing emission which has a SNR of over 3 or 2 in $^{13}$CO or C$^{18}$O respectively, but is not assigned to SNR $>5$ clump are not included in the integrated emission of Fig. \ref{fig:channelmaps} or \ref{fig:integlv} either. There is much more emission visible in the $^{13}$CO images due to the higher abundance of $^{13}$CO relative to C$^{18}$O. The brightest regions in the survey are some of the most massive star-forming regions in the Galaxy, and the W43 and W49A complexes are clearly visible at $l = 30 \fdg 7$ and $43 \fdg 1$, respectively. In the C$^{18}$O maps of Fig. \ref{fig:integlb}, there are a small number of places where noise features have been extracted by FellWalker; for example there are such noise features at $l=43 \fdg 3, \ b=-0 \fdg 01$ and $l=36 \fdg 5, \ b=-0 \fdg 035$. These appear due to the lowering of the detection threshold to SNR $>2$ for the C$^{18}$O data, which was necessary to enable the fainter emission to be seen, but real clumps also emerge which were not visible using the same detection limits as for the $^{13}$CO. Fig. \ref{fig:channelmaps} shows the $^{13}$CO emission integrated over 30\kms\ velocity windows, allowing emission features to be separated along the line of sight and fainter clouds to become more visible than in Fig. \ref{fig:integlb}.

\subsection{Integrated position-velocity maps}\label{sec:emisisonlv}

\begin{figure*}
\hspace*{-1.9cm}
\centering
\includegraphics[width=1.18\textwidth]{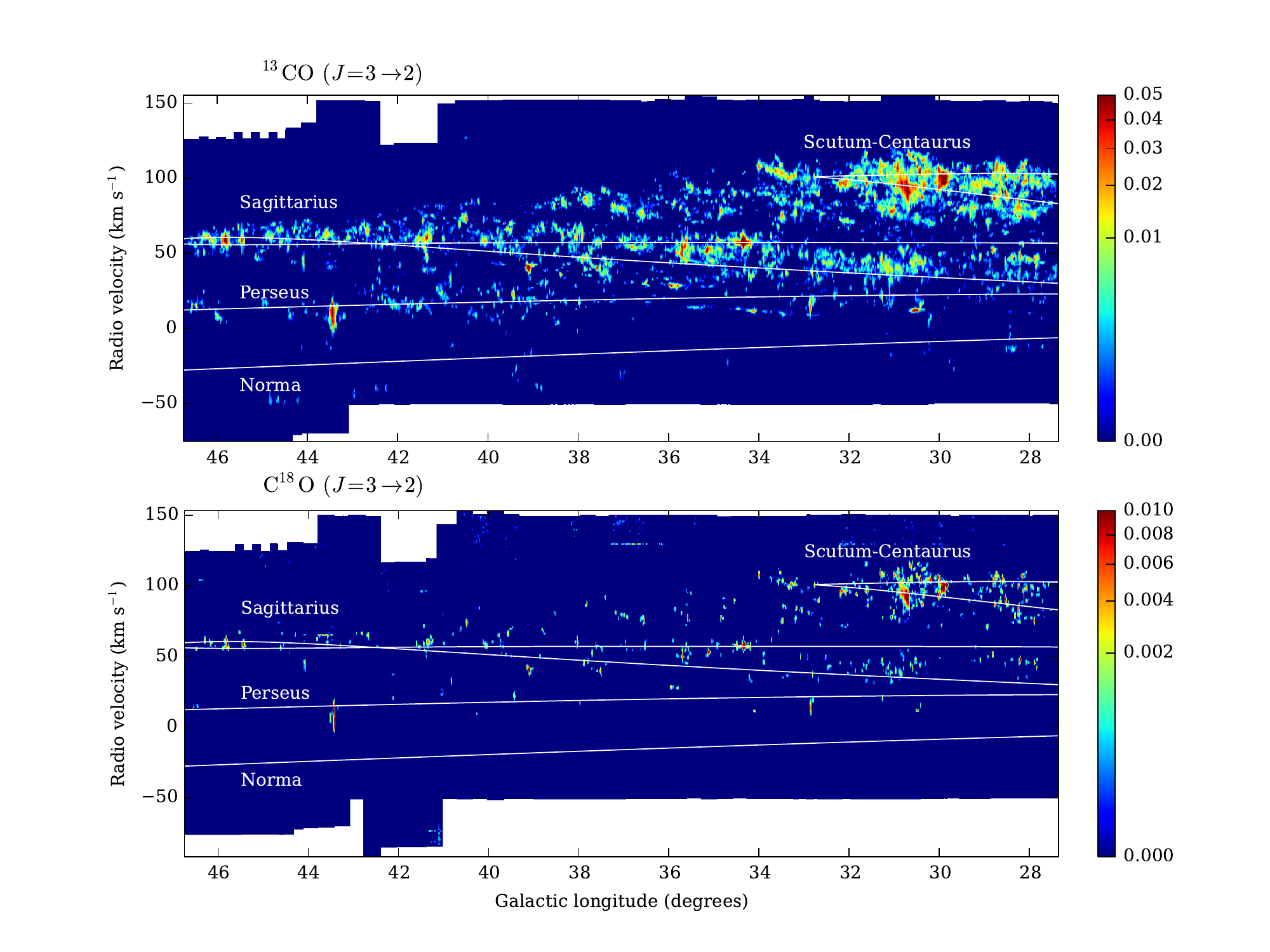}
\caption{Position--velocity diagrams for the $^{13}$CO and C$^{18}$O emission (\tast) with an SNR of at least 3 in CHIMPS in which the emission was integrated over the latitude axis. The colour mapping uses a third-root intensity scale, and has units of K degrees. Each pixel in the longitude axis is the sum of 10 pixels at the same velocity. The overlaid white lines are the spiral arm loci of the four-arm model of \citet{Taylor+Cordes93}, updated in \citet{Cordes04}, projected into the longitude--velocity plane.} \label{fig:integlv}
\end{figure*}

Fig. \ref{fig:integlv} shows the position--velocity diagrams for the $^{13}$CO and C$^{18}$O emission, integrated over the latitude axis. The spiral arms are clearly visible in the $^{13}$CO map as continuous streams of emission, with inter-arm regions also visible as relatively emission-free regions separating the arms. Spiral arms have been overlaid which derive from the models of \citet{Taylor+Cordes93} and \citet{Cordes04}, with the position--velocity--space projections calculated in \citet{Urquhart+13b}. The molecular gas traced by CHIMPS fits reasonably well with this four-arm model, though there are some significant deviations. There is little emission visible which falls on the locus of the distant Norma arm, though a shift of 10--20\kms\ towards negative velocities across the CHIMPS region would be consistent with a number of emission features visible here. 

There is a significant quantity of emission lying between the Scutum--Centaurus and Sagittarius arms, which has been seen before in \cogrs\ \citep{Lee+01,Stark+Lee06}, though not with this clarity. The structure of this emission is much clearer in CHIMPS than in \citet{Dame+01}, \citet{Lee+01}, GRS, or COHRS and has a number of possible explanations. First, this emission could be a minor spiral arm which lies in-between the Scutum--Centaurus and Sagittarius spiral arms. This is suggested by a potential loop feature that extends from the low-longitude end of the survey up to a tangent at approximately $l = 39\degr$, spanning approximately $60-90$\kms\ in velocity. Secondly, this could be an extension of the Scutum--Centaurus arm itself, with an elongated tangent region reaching up to roughly $ 39\degr$ in longitude. Thirdly, this could be a bridging structure of the kind described by \citet{Stark+Lee06} or some similar spur structure, which does not extend far enough to be considered an arm in its own right. Finally, it is possible that this region contains a number of spurs which form their own coherent structures in this parameter space, and which generally extend for several degrees. These coherent objects in position--velocity space might also be one origin of filaments \citep[see][]{Ragan+14}, and arise through the shear of dense regions due to Galactic rotation in the simulations of \citet{Dobbs15}. Tests to distinguish between these scenarios are regrettably beyond the scope of this paper.

Emission in the C$^{18}$O map is much more sparse, though the broad emission from W49A is a prominent feature, and its compact size makes it stand out when compared to the other bright regions such as W43. W49A contains a cluster of ultra-compact \ion{H}{ii} regions \citep{Urquhart+13b}, with powerful H$_{2}$O maser outflows \citep{Smith+09} and strong bipolar outflows seen in \co\ $(J=1\rightarrow 0)$ \citep{Scoville+86}. There are a small number of noise features also visible in the C$^{18}$O map, which are usually easy to identify as they tend to appear at the low- or high-velocity ends of the spectral band. An example of such a noise feature can be seen extending from $\sim 36\degr$ to $37 \fdg 5$ at $\sim 130$\kms.

\section{Data access}\label{sec:access}

The CHIMPS data are available to download from the CANFAR archive\footnote{http://dx.doi.org/10.11570/16.0001}. The data are presented in the FITS format and are available primarily as mosaics which each make up approximately 1 square degree, available at intervals of half a degree. In addition to these mosaics, the individual cubes which each represent a single observation (or several observations for the co-added cubes) are available, along with the variance arrays for the mosaics and individual cubes. The integrated emission maps in $l-b$ and $l-v$ space of Section \ref{sec:emissionmaps} can also be downloaded. The data are presented in \tast\, with data cubes in units of K, and the integrated $l$--$b$ and $l$--$v$ maps have units of K$\,$degrees and \kkms\ respectively.

The raw data can be downloaded from the Canadian Astronomy Data Centre's JCMT Science Archive using the Project IDs listed in Section \ref{sec:obs}.

\section{Comparison with GRS and COHRS}\label{sec:comparison}

The GRS mapped the inner Galactic plane in $^{13}$CO $(J = 1 \rightarrow 0)$ at an angular resolution approximately three times lower than CHIMPS. Since the critical density of the $J = 1 \rightarrow 0$ transition is also lower than that of $J = 3 \rightarrow 2$ ($\sim 10^{3} \, \mathrm{cm}^{-3}$ and $\sim 10^{4} \, \mathrm{cm}^{-3}$ at 10\,K, respectively), the molecular gas traced by CHIMPS is much more concentrated spatially (and presumably traces higher column densities) than in GRS, allowing us to see the dense cores and filaments which appear to be almost ubiquitous and closely associated with the star formation seen in continuum surveys such as Hi-GAL \citep{Molinari+10b}.

\begin{figure*}
\centering
\includegraphics[width=0.9\textwidth]{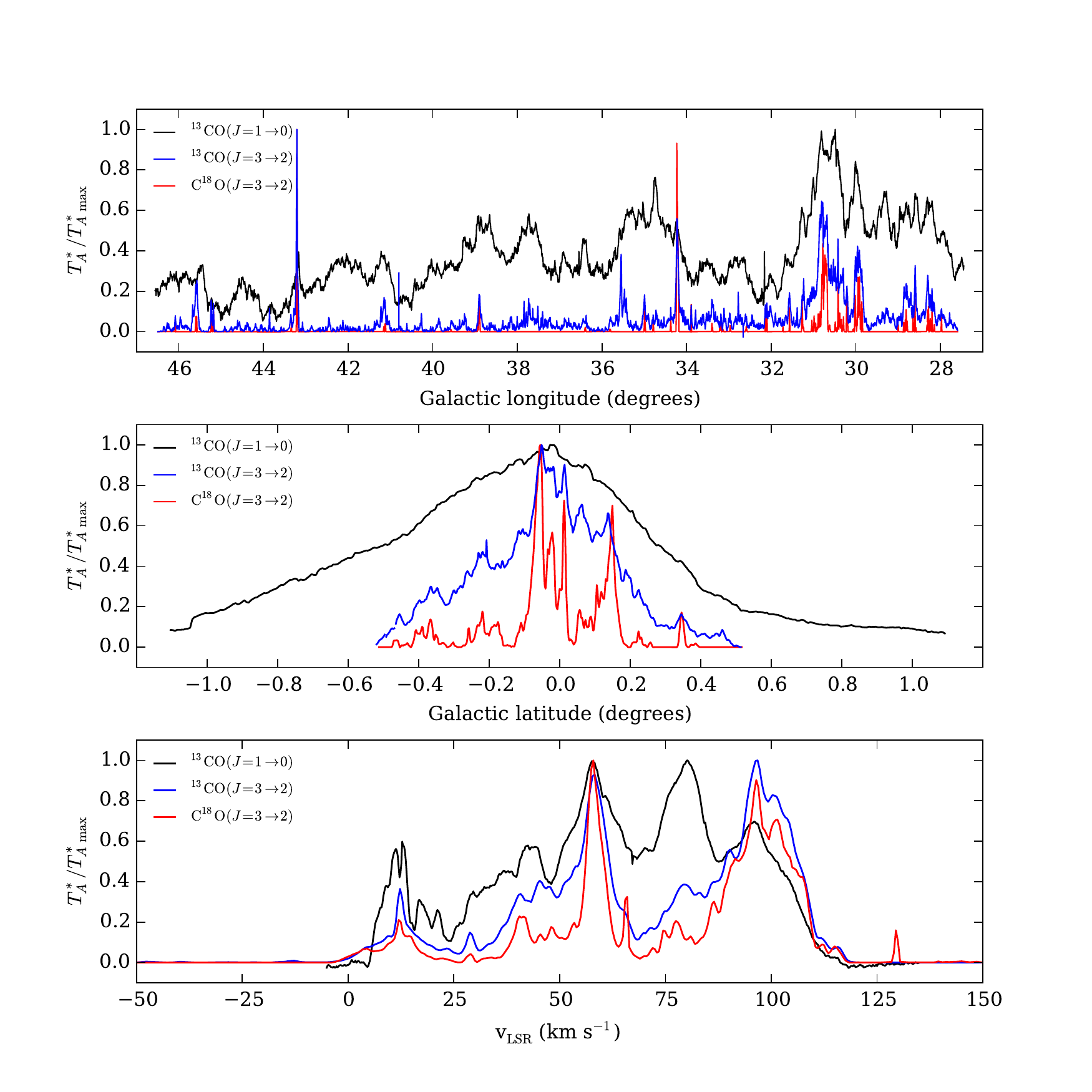}
\caption{Integrated (one-dimensional) longitudinal, latitudinal, and velocity profiles for the GRS and the two CHIMPS isotopologues. In each case, the one-dimensional profile was created by integrating over the two orthogonal axes. The \tast\ intensity is normalized to the peak intensity in the profile.} \label{fig:profiles}
\end{figure*}

Fig. \ref{fig:profiles} shows the integrated intensity $l$, $b$ and $v$ profiles for the GRS over the extent of the CHIMPS region, and of the two CHIMPS tracers. In each case, the profiles show the intensity normalised to the peak intensity in the profile and integrated over both orthogonal axes. For the two CHIMPS tracers, the extracted emission described in Section \ref{sec:emissionlb} was used to make the profiles, whereas the GRS data were integrated over all velocity channels. In the longitudinal profile (integrated over latitude and velocity), we find that the regions of strongest emission in the GRS are generally coincident with a peak in the CHIMPS data, though the \cob\ only appears at the highest column density regions. The peak in the longitudinal profile at $l \approx 34.2\degr$, for example, is much more sharply peaked in \cob\ than \coa\ which is itself more sharply peaked than the \cogrs, possibly indicating self-absorption in the $^{13}$CO spectrum, or greater turbulence in the lower density material. Additionally, the star-forming region W49A located at $l \approx 43 \fdg 2$ stands out with a strong, sharp peak in \coa. 

The latitudinal profiles (integrated over longitude and velocity) also display a trend of increasing sharpness as we move into denser gas tracers and at higher resolution as expected, and the normalised intensity of \coa\ is close to zero at the limits of the survey. It is therefore reasonable to suggest that our latitude range for CHIMPS is not missing significant quantities of emission in the inner Galactic plane. The two $^{13}$CO transitions have profiles which are asymmetric about $b=0\degr$ which can be attributed to both the warp in the Milky Way's disc, and a parallax effect caused by the position of the Sun between 4 and 30 pc above the Galactic plane \citep{deVaucouleurs+Malik69,Stenholm75,Bahcall+Bahcall85}. 

The velocity profiles (integrated over longitude and latitude) are again more sharply peaked in the CHIMPS tracers compared to GRS as the diffuse gas component becomes transparent, leaving the distributions of gas denser than $10^{4}$\,cm$^{-3}$. The C$^{18}$O peak at $\approx 130$\kms\ which is not seen in the other tracers is caused by noise artefacts that appear as a result of the less stringent noise criteria applied to this isotopologue described in Section \ref{sec:extraction}.

In comparison to COHRS, a JCMT survey of \cohrs\ covering much of the CHIMPS area, there is significantly less faint and extended emission in the CHIMPS data. The higher optical depths and self-absorption in the \co\ data suppress the emission peaks and there is an additional effect of photon pumping at high optical depths which reduces the effective critical density of \cohrs, enhancing emission from more diffuse gas. These effects combine to reduce the contrast between the between high- and low-column density regions in the COHRS data. There is, therefore, more contrast between the faint and bright emission in CHIMPS and massive cores appear to have a steeper density profile as more of the densest gas can be observed. This means that it is possible to deduce dense gas masses in CHIMPS with improved accuracy, and the sensitivity in terms of column density is less complex due to the lesser contribution of photon pumping.

\begin{figure*}
\centering
\includegraphics[width=0.8\textwidth]{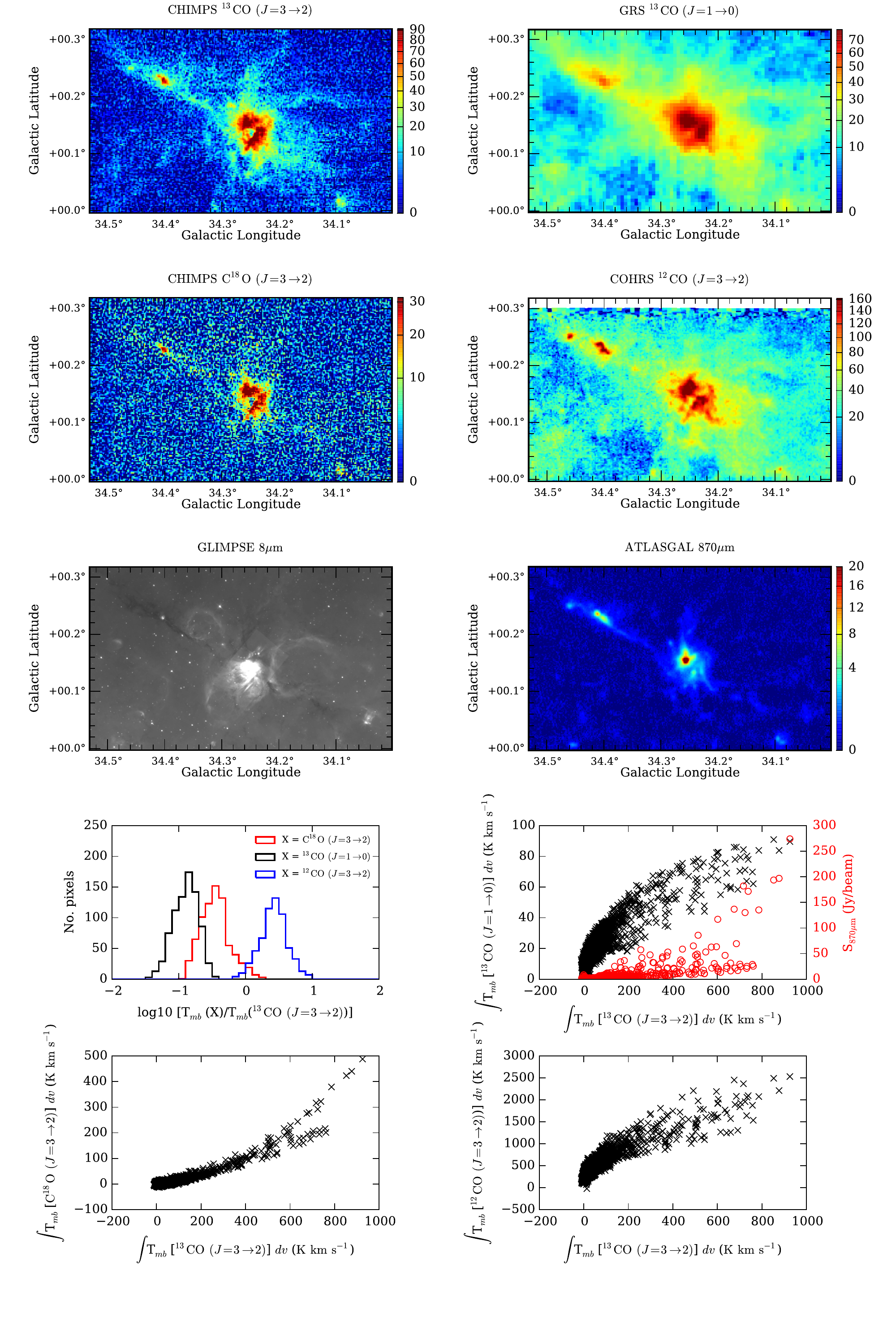}
\caption{Top: images of the G34.2+0.1 region in the two CHIMPS isotopologues, and imaging from GRS, COHRS, GLIMPSE and ATLASGAL. The images are shown in their native resolution, and the CHIMPS, COHRS and GRS images are integrated over 45--70\kms. The units on the integrated \tmb\ intensity scales are \kkms, with the exception of the ATLASGAL image, which is in units of Jy per beam. A square-root scaling is used in each image. Bottom: histogram of the intensity ratios of the different species compared to \coa\ calculated on a voxel-by-voxel basis for all voxels brighter than 5$\sigma$, and pixel-by-pixel correlations for all pixels in the integrated images.} \label{fig:G34}
\end{figure*}

A region centred on Galactic coordinates $l = 34 \fdg 25, b=+0 \fdg 15$ and with the velocity range $v_\mathrm{lsr} = 45 - 70$\kms\ (hereafter the `G34 region', also known by the identifier IRAS 18507+0110), which contains a number of ultra-compact \ion{H}{ii} regions seen in the Red MSX Source survey \citep{Lumsden+13}, is shown in Fig. \ref{fig:G34}. This region lies at a distance of 4.0 kpc based on the water maser parallax measurements of G34.26+0.15 \citep{Hofner+Churchwell96}, and has a Galactocentric distance of $\sim 4.5\,$kpc, based on the Galactic rotation curve of \citet{Brand+Blitz93} and central velocity of 57.5 \kms. CHIMPS, COHRS and ATLASGAL ($870\,\mu$m) imaging have been smoothed spatially using Gaussian kernels with FWHM of 43.4, 42.9 and 41.8 arcsec respectively in order to match the 46 arcsec resolution of the GRS and re-gridded to the GRS pixel size. Intensity scales in the various CO data were converted from \tast\ to \tmb\ by dividing by main beam efficiencies of $\eta_\mathrm{mb} = 0.72$ and 0.61 for CHIMPS and COHRS, respectively \citep{Buckle+09}, and $\eta_\mathrm{mb} = 0.48$ for GRS \citep{Jackson+06}. 

The various CO cubes were aligned in three dimensions, and we present histograms (left column, second row from bottom) of the voxel-by-voxel intensity ratios of $^{13}$CO$ (J = 1 \rightarrow 0)$,  $^{12}$CO$ (J = 3 \rightarrow 2)$, and \cob\ with respect to \coa. The intensity ratio was measured only for voxels in which both species have an intensity above five times the rms value of all voxels each cube. In instances where both species are optically thin, the intensity ratio ought to be equal to the abundance ratio of the species. It is unlikely, however, that a significant number of voxels are optically thin in both species for any pairing. The black histogram, showing the intensity ratio distribution of  $^{13}$CO$ (J = 1 \rightarrow 0)$ to \coa, has a median value close to 0.1. For optically thin gas at temperatures significantly greater than $h\nu/k$, we should expect this ratio to approach a value of one ninth since $T_R (J+1\rightarrow J)/T_R (J \rightarrow J-1) = (J+1/J)^2$. Deviations from small $\tau$ in either transition, along with uncertainties in the intensity measurement contribute towards broadening this distribution. 

The red histogram shows the intensity ratio of the two CHIMPS isotopologues, \cob\ to \coa, and in the cases where both voxels are optically thin, we would expect to recover the abundance ratio of C$^{18}$O to $^{13}$CO. At a Galactocentric distance of 4.5$\,$kpc, the isotopic abundance ratios for $^{12}$C/$^{13}$C \citep{Milam+05} and $^{16}$O/$^{18}$O \citep{Wilson+Rood94} indicate that we should expect an abundance ratio of C$^{18}$O/$^{13}$CO $\sim 1/6$, which is consistent with these measurements. The blue histogram which measures the intensity ratio of $^{12}$CO$ (J = 3 \rightarrow 2)$ to \coa\ has a median value of $< 10$, whereas the \citet{Milam+05} relation predicts a value close to 50. It is unlikely that any optically thin $^{12}$CO$ (J = 3 \rightarrow 2)$ emission is detected where \coa\ is also recovered and so the intensity ratio is suppressed, and further reduced by self-absorption which is likely to be significant in this high optical depth transition.

The pixel-to-pixel correlations of $^{13}$CO$ (J = 1 \rightarrow 0)$,  $^{12}$CO$ (J = 3 \rightarrow 2)$, \cob\ and $870\,\mu$m with \coa\ of the integrated images for the G34 region are also presented in Fig. \ref{fig:G34}. In the correlations between the different CO isotopologues there are strong optical depth effects visible where the denser tracer dominates in the brightest regions, and these effects are more significant in the integrated image, where any optically thin voxels are folded into an optically thick column. These distributions also contain noise pixels, though these are not significant when integrated over the velocity range, and make up the high concentration of points towards the origin. The correlation between $870\,\mu$m and \coa\ emission was measured only for pixels with intensities greater than five times the rms $870\,\mu$m value. For the majority of eligible pixels a linear correlation is visible between dust and CO emission, extending from $\sim 0$ to $800\,$\kkms\ in \coa\ and $\sim 0$ to $50\,$Jy in S$_{870}$, but there are a number of pixels in which the dust emission becomes significantly brighter. This could be caused by \coa\ emission becoming optically thick where the brightest $870\,\mu$m emission is, though these may also correspond to a small number of objects that are bright and compact in the continuum data but disappear into the background in degraded-resolution \coa\ image.

This study of the G34 region shows that the brightness temperatures measured within the CHIMPS data are consistent with comparable survey data, and demonstrate that they, when used in conjunction with data sets such as GRS and COHRS, provide a more complete picture.

\section{Example CHIMPS data}\label{sec:previews}

\begin{landscape}
\begin{figure}
\centering
\includegraphics[width=1.3\textwidth]{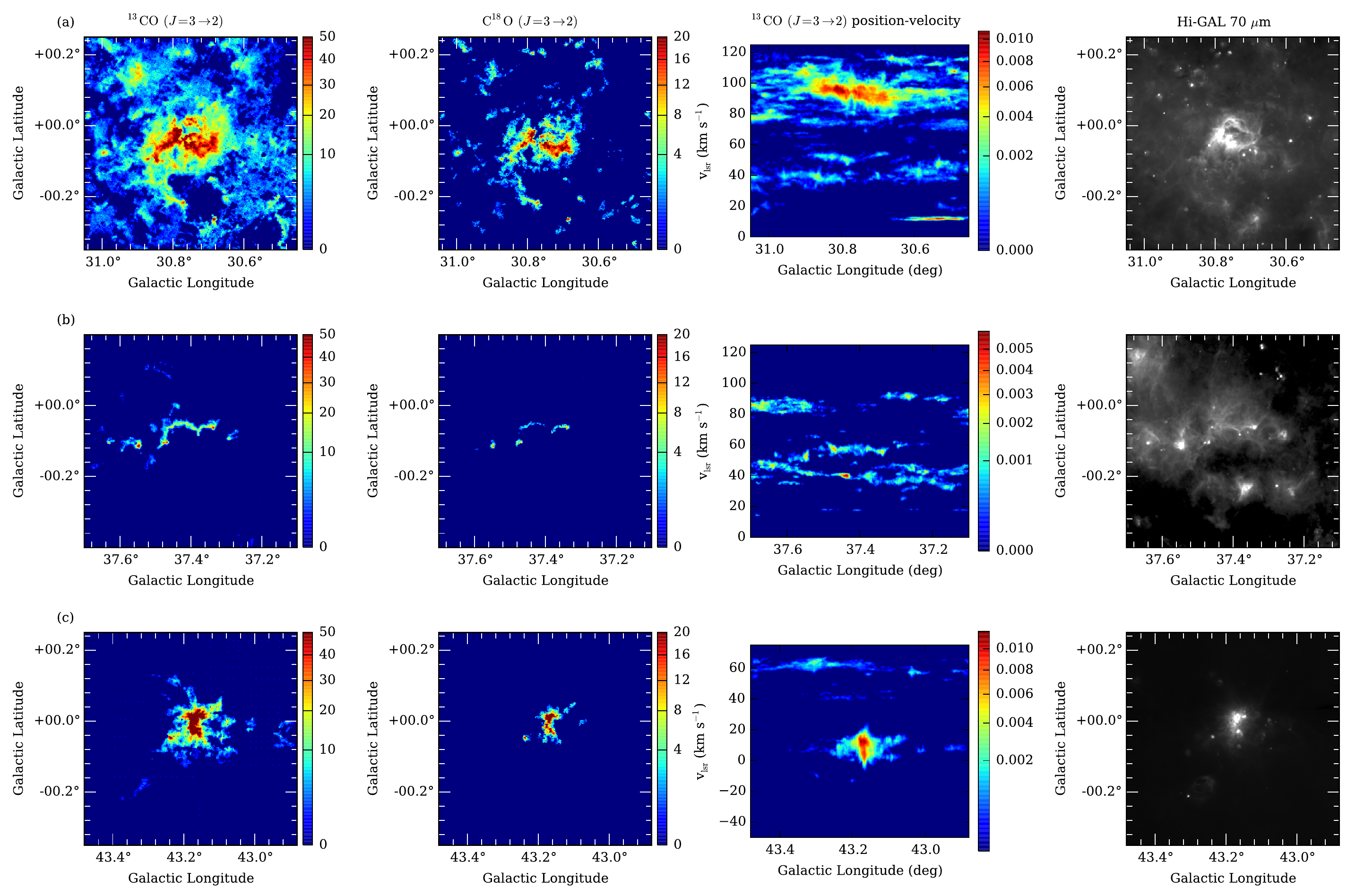}
\caption{Close-ups of several interesting regions in the CHIMPS data: a) The W43 star-forming complex, centred on $l = 30 \fdg 75, b=-0 \fdg 05$ and integrated over 80--110 km s$^{-1}$;  b) A filamentary structure centred on $l = 37 \fdg 4, b=-0 \fdg 1$ and integrated over 50--65 km s$^{-1}$; c) W49A star-forming complex centred on $l = 43 \fdg 18, b=-0 \fdg 05$ and integrated over $-$30 to 30 km s$^{-1}$. The first two columns from the left-hand side show the integrated intensity in \coa\ and \cob\ respectively, while the third from left column shows the region integrated over latitude, and the fourth column shows the 70 $\umu$m images of regions from Hi-GAL. In the integrated images in the first and second columns, the intensity scale shows the integrated corrected antenna temperature \tast\ in units of K\,\kms, and the integrated \tast\ intensity scale of the position--velocity maps in the third column has units of K\,degrees.} \label{fig:previews}
\end{figure}
\end{landscape}

Some sample close-ups of the CHIMPS data are illustrated in Fig. \ref{fig:previews} in which we examine some of the interesting regions in the survey in integrated \coa\ (first from left column), \cob\ (second from left column) maps, \coa\  position-velocity space (second from right column) and the 70 $\umu$m Hi-GAL images for comparison (right column). In each case, the CHIMPS data used contain the emission extracted according to Section \ref{sec:extraction}.

The W43 star-forming region ($l = 30 \fdg 75, b = -0 \fdg 05$), illustrated in row (a) of Fig. \ref{fig:previews} is the most prominent region within CHIMPS (see Fig. \ref{fig:integlb}). At a distance of 5.5 kpc \citep{Zhang+14}, W43 lies at the tangent of the Scutum--Centaurus arm and its meeting point with the near-end of the Long Bar \citep{Nguyen-Luong+11}. The region presented has been integrated over a velocity range of $v_{\mathrm{lsr}} = 80-110$\kms, identified by \citet{Nguyen-Luong+11} as the central part of the cloud; indeed, this velocity range is extremely well matched by the CHIMPS spectra of the region in both \coa\ and \cob. W43 is frequently referred to as a `mini-starburst' \citep[e.g.][]{Motte+03}, implying a high star formation efficiency, and while it contains a high--column--density ridge known as `MM1' with a high star-forming efficiency of 8\% \citep{Louvet+14}, the region as a whole does not appear be particularly unusual.

\citet{Eden+12} find that, while the fraction of mass in dense BGPS clumps compared to the mass in \cogrs\ for clumps coincident with the W43 \ion{H}{ii} region is high, the median value of this quantity for all clumps in the region is not enhanced when compared to other regions along the same line of sight. Similarly \citet{Moore+12} and \citet{Eden+15} find that the star-formation efficiency averaged at this Galactocentric radius is also unexceptional, with all three studies suggesting that W43 is consistent with being part of a normal distribution of star-forming properties. W43 is also the subject of the recent pilot study for The \ion{H}{i}, OH, Recombination Line Survey of the Milky Way \citep[THOR;][]{Bihr+15}, who revise the mass in \ion{H}{i} of the complex, finding a lower limit of $6.6 \times 10^{6}$ M$_{\sun}$.

A striking filament visible in the CHIMPS data (see Fig. \ref{fig:integlb}) is examined in row (b) of Fig. \ref{fig:previews}, centred on $l = 37 \fdg 4$, $b=-0 \fdg 1$  and integrated over 50--65\kms. The structure is coherent in position--velocity space, and so may be viewed as a single structure. With a single high-density ridge and little or no diffuse gas surrounding it, this filament appears to have an especially compact profile. The relatively low contrast of the filament in the 70 $\umu$m image compared to the molecular gas images suggests that the filament is largely cool, though the peaks at either end of the filament are associated with \ion{H}{ii} regions, such as HRDS G037.468--0.105 \citep{Bania+12} and IRAS 18571+0349 \citep{Johnston+09}, and several sites of massive star formation are also present. The filament lies at a distance of 9.6 kpc \citep{Bania+12}, assuming that the coherence in position--velocity space implies a single distance, and contains a string of 1.1 mm clumps identified in the BGPS \citep{Rosolowsky+10}. The total length of the filament, following its shape, is approximately 14 arcmin, which corresponds to a length of $30\,$pc. Its width, which remains roughly constant across its length, is $\sim$22 arcsec corresponding to $\sim 1\,$pc and implying an average aspect ratio of about 30.

Further comparison between the \coa\ and $70\,\mu$m images reveal diffuse material that appears to be missing in CHIMPS, and there are also several compact sources in the 70 $\umu$m image which do not appear in the integrated position--position maps, but appear at different velocities in position--velocity space. The $70\,\mu$m image would appear to show that the filament lies at the intersection of a small number of bubble edges, and when viewed alongside the CHIMPS data, there is a suggestion that these bubbles are sweeping up a significant quantity of gas culminating in the dense ridge of this filament.


Row (c) of Fig. \ref{fig:previews} is centred on  $l = 43 \fdg 18$, $b=-0 \fdg 05$ and shows the massive star-forming region W49 integrated over the velocity range of $-30$ to 30\kms. W49, located at a distance of 11.11 kpc \citep{Zhang+13}, is associated with the brightest emission visible in CHIMPS and is the most luminous \citep[$\sim 10^{7}\,\mathrm{L}_{\sun}$][]{Sievers+91} star-forming region, and one of the most massive in the Galaxy \citep[$M_{\mathrm{gas}} \sim 1.1 \times 10^{6} M_{\sun}$;][]{Galvan-Madrid+13}. The emission in W49 is extremely broad, spanning $\approx 35$\kms\ in \coa\ traced by CHIMPS owing to high velocity bipolar outflows \citep{Scoville+86}, and its three-dimensional structure can be described as a hub-filament system, with filaments converging on the W49N and W49S clusters. The whole complex is thought to contain $\sim 13$ ultra-compact \ion{H}{ii} regions \citep{Urquhart+13b}, and contains sufficient mass to form several young massive clusters. \citet{Moore+12} find that the star formation in W49, which has a flatter-than-normal luminosity function, is exceptional in terms of its efficiency, and as such may be considered truly starburst-like.

\section{Summary}\label{sec:summary}

We present data from CHIMPS, a survey of the $J=3 \rightarrow 2$ rotational transition of $^{13}$CO and C$^{18}$O in a region of the inner Galactic plane, spanning approximately $28\degr \leq l \leq 46\degr$ and $|b| \leq 0 \fdg 5$, which is now publicly available at http://dx.doi.org/10.11570/16.0001. The data have a spatial resolution of 15 arcsec and a spectral resolution of 0.5\kms\ per channel, with a bandwidth of 200\kms. With a median rms of $\sim$ 0.6\,K in the \coa\ spectra at this resolution, the sensitivity corresponds to a column density of roughly $N_{^{13}\mathrm{CO}} \sim 3 \times 10^{14}\,$cm$^{-2}$ or $N_{\mathrm{H}_2} \sim 3 \times 10^{20}\,$cm$^{-2}$ for optically thin gas at an excitation temperature of 10 K. The \cob\ spectra have a median rms of $0.7\,$K per channel, corresponding to $N_{\mathrm{C}^{18}\mathrm{O}} \sim 4 \times 10^{14}\,$cm$^{-2}$ or $N_{\mathrm{H}_2} \sim 4 \times 10^{21}\,$cm$^{-2}$.

The relatively low abundances of the two CHIMPS isotopologues compared to \co\ (the relative abundances of  $^{13}$CO and C$^{18}$O  compared to \co\ are $\sim 10^{-2}$ and $\sim 10^{-3}$, respectively) mean that they become optically thick at much higher column densities. The CHIMPS data, therefore, may serve as an excellent resource for finding the dense substructures in molecular clouds that fuel star formation \citep{Molinari+10b}. When used in conjunction with other CO surveys such as COHRS \citep{Dempsey+13} and GRS \citep{Jackson+06} which trace different gas components, these data will help to provide a clearer picture of star-formation, and molecular gas dynamics. CHIMPS also complements the wealth of submillimetre surveys such as the JPS, Hi-GAL, ATLASGAL, BGPS and infra-red surveys like \textit{WISE}, GLIMPSE and MIPSGAL. 

The combination of the resolution and the high gas critical densities of these observations means that CHIMPS shows a high contrast between high- and low- column density regions, and therefore serves as an ideal resource for studying Galactic structure. In position-velocity space, the integrated emission from CHIMPS displays a significant quantity of complex emission between the Sagittarius and Scutum-Centaurus arms, possibly indicating a bridge structure, or a minor spiral arm, or a spur or series of spurs.

These data have been used in the studies of \citet{Moore+15}, \citet{Urquhart+13b}, and \citet{Eden+12,Eden+13}. A catalogue of molecular clouds found in CHIMPS will be released in an upcoming paper (Rigby et al., in preparation) with a study of the masses of molecular cloud structures across the survey area.

\section*{Acknowledgements}

AJR would like to thank David Berry and especially Malcolm Currie for their assistance with {\sc orac-dr} and other Starlink applications. AJR also acknowledges the support of an STFC-funded studentship, and the support of the Royal Astronomical Society for overseas travel to present CHIMPS data. The James Clerk Maxwell Telescope has historically been operated by the Joint Astronomy Centre on behalf of the Science and Technology Facilities Council of the United Kingdom, the National Research Council of Canada and the Netherlands Organisation for Scientific Research. This research used the facilities of the Canadian Astronomy Data Centre operated by the National Research Council of Canada with the support of the Canadian Space Agency. This research made use of the NASA Astrophysical Data System. This research has also made use of the SIMBAD data base, operated at CDS, Strasbourg, France. This research also made use of {\sc astropy}, a community-developed core {\sc python} package for Astronomy \citep{astropy}. This research made use of {\sc aplpy}, an open-source plotting package for {\sc python} hosted at http://aplpy.github.com. MZ acknowledges support from the China Ministry of Science and Technology under the State Key Development Program for Basic Research (2012CB821800).



\bibliographystyle{mnras}
\bibliography{AJR_References_CHIMPS1}

\begin{thebibliography}{}
\makeatletter
\relax
\def\mn@urlcharsother{\let\do\@makeother \do\$\do\&\do\#\do\^\do\_\do\%\do\~}
\def\mn@doi{\begingroup\mn@urlcharsother \@ifnextchar [ {\mn@doi@}
  {\mn@doi@[]}}
\def\mn@doi@[#1]#2{\def\@tempa{#1}\ifx\@tempa\@empty \href
  {http://dx.doi.org/#2} {doi:#2}\else \href {http://dx.doi.org/#2} {#1}\fi
  \endgroup}
\def\mn@eprint#1#2{\mn@eprint@#1:#2::\@nil}
\def\mn@eprint@arXiv#1{\href {http://arxiv.org/abs/#1} {{\tt arXiv:#1}}}
\def\mn@eprint@dblp#1{\href {http://dblp.uni-trier.de/rec/bibtex/#1.xml}
  {dblp:#1}}
\def\mn@eprint@#1:#2:#3:#4\@nil{\def\@tempa {#1}\def\@tempb {#2}\def\@tempc
  {#3}\ifx \@tempc \@empty \let \@tempc \@tempb \let \@tempb \@tempa \fi \ifx
  \@tempb \@empty \def\@tempb {arXiv}\fi \@ifundefined
  {mn@eprint@\@tempb}{\@tempb:\@tempc}{\expandafter \expandafter \csname
  mn@eprint@\@tempb\endcsname \expandafter{\@tempc}}}

\bibitem[\protect\citeauthoryear{{Aguirre} et~al.,}{{Aguirre}
  et~al.}{2011}]{Aguirre+11}
{Aguirre} J.~E.,  et~al., 2011, \mn@doi [\apjs] {10.1088/0067-0049/192/1/4},
  \href {http://adsabs.harvard.edu/abs/2011ApJS..192....4A} {192, 4}

\bibitem[\protect\citeauthoryear{{Astropy Collaboration} et~al.,}{{Astropy
  Collaboration} et~al.}{2013}]{astropy}
{Astropy Collaboration} et~al., 2013, \mn@doi [\aap]
  {10.1051/0004-6361/201322068}, \href
  {http://adsabs.harvard.edu/abs/2013A%26A...558A..33A} {558, A33}

\bibitem[\protect\citeauthoryear{{Bahcall} \& {Bahcall}}{{Bahcall} \&
  {Bahcall}}{1985}]{Bahcall+Bahcall85}
{Bahcall} J.~N.,  {Bahcall} S.,  1985, \mn@doi [\nat] {10.1038/316706a0}, \href
  {http://adsabs.harvard.edu/abs/1985Natur.316..706B} {316, 706}

\bibitem[\protect\citeauthoryear{{Bania}, {Anderson}  \& {Balser}}{{Bania}
  et~al.}{2012}]{Bania+12}
{Bania} T.~M.,  {Anderson} L.~D.,   {Balser} D.~S.,  2012, \mn@doi [\apj]
  {10.1088/0004-637X/759/2/96}, \href
  {http://adsabs.harvard.edu/abs/2012ApJ...759...96B} {759, 96}

\bibitem[\protect\citeauthoryear{{Benjamin} et~al.,}{{Benjamin}
  et~al.}{2003}]{Benjamin+03}
{Benjamin} R.~A.,  et~al., 2003, \mn@doi [\pasp] {10.1086/376696}, \href
  {http://adsabs.harvard.edu/abs/2003PASP..115..953B} {115, 953}

\bibitem[\protect\citeauthoryear{{Berry}}{{Berry}}{2015}]{Berry15}
{Berry} D.~S.,  2015, \mn@doi [Astronomy and Computing]
  {10.1016/j.ascom.2014.11.004}, \href
  {http://adsabs.harvard.edu/abs/2015A%26C....10...22B} {10, 22}

\bibitem[\protect\citeauthoryear{{Berry}, {Reinhold}, {Jenness}  \&
  {Economou}}{{Berry} et~al.}{2007}]{Berry+07}
{Berry} D.~S.,  {Reinhold} K.,  {Jenness} T.,   {Economou} F.,  2007, in {Shaw}
  R.~A.,  {Hill} F.,   {Bell} D.~J.,  eds,  Astronomical Society of the Pacific
  Conference Series Vol. 376, Astronomical Data Analysis Software and Systems
  XVI. p.~425

\bibitem[\protect\citeauthoryear{{Bihr} et~al.,}{{Bihr} et~al.}{2015}]{Bihr+15}
{Bihr} S.,  et~al., 2015, preprint, \href
  {http://adsabs.harvard.edu/abs/2015arXiv150505176B} {} (\mn@eprint {arXiv}
  {1505.05176})

\bibitem[\protect\citeauthoryear{{Bolatto}, {Wolfire}  \& {Leroy}}{{Bolatto}
  et~al.}{2013}]{Bolatto+13}
{Bolatto} A.~D.,  {Wolfire} M.,   {Leroy} A.~K.,  2013, \mn@doi [\araa]
  {10.1146/annurev-astro-082812-140944}, \href
  {http://adsabs.harvard.edu/abs/2013ARA%26A..51..207B} {51, 207}

\bibitem[\protect\citeauthoryear{{Brand} \& {Blitz}}{{Brand} \&
  {Blitz}}{1993}]{Brand+Blitz93}
{Brand} J.,  {Blitz} L.,  1993, \aap, \href
  {http://adsabs.harvard.edu/abs/1993A%26A...275...67B} {275, 67}

\bibitem[\protect\citeauthoryear{{Buckle} et~al.,}{{Buckle}
  et~al.}{2009}]{Buckle+09}
{Buckle} J.~V.,  et~al., 2009, \mn@doi [\mnras]
  {10.1111/j.1365-2966.2009.15347.x}, \href
  {http://cdsads.u-strasbg.fr/abs/2009MNRAS.399.1026B} {399, 1026}

\bibitem[\protect\citeauthoryear{{Carey} et~al.,}{{Carey}
  et~al.}{2009}]{Carey+09}
{Carey} S.~J.,  et~al., 2009, \mn@doi [\pasp] {10.1086/596581}, \href
  {http://adsabs.harvard.edu/abs/2009PASP..121...76C} {121, 76}

\bibitem[\protect\citeauthoryear{{Churchwell} et~al.,}{{Churchwell}
  et~al.}{2009}]{Churchwell+09}
{Churchwell} E.,  et~al., 2009, \mn@doi [\pasp] {10.1086/597811}, \href
  {http://adsabs.harvard.edu/abs/2009PASP..121..213C} {121, 213}

\bibitem[\protect\citeauthoryear{{Cordes}}{{Cordes}}{2004}]{Cordes04}
{Cordes} J.~M.,  2004, in {Clemens} D.,  {Shah} R.,   {Brainerd} T.,  eds,
  Astronomical Society of the Pacific Conference Series Vol. 317, Milky Way
  Surveys: The Structure and Evolution of our Galaxy. p.~211

\bibitem[\protect\citeauthoryear{{Currie}}{{Currie}}{2013}]{Currie13}
{Currie} M.~J.,  2013, in {Friedel} D.~N.,  ed.,  Astronomical Society of the
  Pacific Conference Series Vol. 475, Astronomical Data Analysis Software and
  Systems XXII. p.~341

\bibitem[\protect\citeauthoryear{{Currie}, {Draper}, {Berry}, {Jenness},
  {Cavanagh}  \& {Economou}}{{Currie} et~al.}{2008}]{Currie+08}
{Currie} M.~J.,  {Draper} P.~W.,  {Berry} D.~S.,  {Jenness} T.,  {Cavanagh} B.,
    {Economou} F.,  2008, in {Argyle} R.~W.,  {Bunclark} P.~S.,   {Lewis}
  J.~R.,  eds,  Astronomical Society of the Pacific Conference Series Vol. 394,
  Astronomical Data Analysis Software and Systems XVII. p.~650

\bibitem[\protect\citeauthoryear{{Currie}, {Berry}, {Jenness}, {Gibb}, {Bell}
  \& {Draper}}{{Currie} et~al.}{2014}]{Currie+14}
{Currie} M.~J.,  {Berry} D.~S.,  {Jenness} T.,  {Gibb} A.~G.,  {Bell} G.~S.,
  {Draper} P.~W.,  2014, in {Manset} N.,  {Forshay} P.,  eds,  Astronomical
  Society of the Pacific Conference Series Vol. 485, Astronomical Data Analysis
  Software and Systems XXIII. p.~391

\bibitem[\protect\citeauthoryear{{Dame}, {Hartmann}  \& {Thaddeus}}{{Dame}
  et~al.}{2001}]{Dame+01}
{Dame} T.~M.,  {Hartmann} D.,   {Thaddeus} P.,  2001, \mn@doi [\apj]
  {10.1086/318388}, \href {http://adsabs.harvard.edu/abs/2001ApJ...547..792D}
  {547, 792}

\bibitem[\protect\citeauthoryear{{Dempsey}, {Thomas}  \& {Currie}}{{Dempsey}
  et~al.}{2013}]{Dempsey+13}
{Dempsey} J.~T.,  {Thomas} H.~S.,   {Currie} M.~J.,  2013, \mn@doi [\apjs]
  {10.1088/0067-0049/209/1/8}, \href
  {http://adsabs.harvard.edu/abs/2013ApJS..209....8D} {209, 8}

\bibitem[\protect\citeauthoryear{{Dobbs}}{{Dobbs}}{2015}]{Dobbs15}
{Dobbs} C.~L.,  2015, \mn@doi [\mnras] {10.1093/mnras/stu2585}, \href
  {http://adsabs.harvard.edu/abs/2015MNRAS.447.3390D} {447, 3390}

\bibitem[\protect\citeauthoryear{{Duarte-Cabral}, {Acreman}, {Dobbs},
  {Mottram}, {Gibson}, {Brunt}  \& {Douglas}}{{Duarte-Cabral}
  et~al.}{2015}]{Duarte-Cabral+15}
{Duarte-Cabral} A.,  {Acreman} D.~M.,  {Dobbs} C.~L.,  {Mottram} J.~C.,
  {Gibson} S.~J.,  {Brunt} C.~M.,   {Douglas} K.~A.,  2015, \mn@doi [\mnras]
  {10.1093/mnras/stu2586}, \href
  {http://adsabs.harvard.edu/abs/2015MNRAS.447.2144D} {447, 2144}

\bibitem[\protect\citeauthoryear{{Eden}, {Moore}, {Plume}  \& {Morgan}}{{Eden}
  et~al.}{2012}]{Eden+12}
{Eden} D.~J.,  {Moore} T.~J.~T.,  {Plume} R.,   {Morgan} L.~K.,  2012, \mn@doi
  [\mnras] {10.1111/j.1365-2966.2012.20840.x}, \href
  {http://adsabs.harvard.edu/abs/2012MNRAS.422.3178E} {422, 3178}

\bibitem[\protect\citeauthoryear{{Eden}, {Moore}, {Morgan}, {Thompson}  \&
  {Urquhart}}{{Eden} et~al.}{2013}]{Eden+13}
{Eden} D.~J.,  {Moore} T.~J.~T.,  {Morgan} L.~K.,  {Thompson} M.~A.,
  {Urquhart} J.~S.,  2013, \mn@doi [\mnras] {10.1093/mnras/stt279}, \href
  {http://adsabs.harvard.edu/abs/2013MNRAS.431.1587E} {431, 1587}

\bibitem[\protect\citeauthoryear{{Eden}, {Moore}, {Urquhart}, {Elia}, {Plume},
  {Rigby}  \& {Thompson}}{{Eden} et~al.}{2015}]{Eden+15}
{Eden} D.~J.,  {Moore} T.~J.~T.,  {Urquhart} J.~S.,  {Elia} D.,  {Plume} R.,
  {Rigby} A.~J.,   {Thompson} M.~A.,  2015, \mn@doi [\mnras]
  {10.1093/mnras/stv1323}, \href
  {http://adsabs.harvard.edu/abs/2015MNRAS.452..289E} {452, 289}

\bibitem[\protect\citeauthoryear{{Elmegreen}}{{Elmegreen}}{1980}]{Elmegreen+80}
{Elmegreen} D.~M.,  1980, \mn@doi [\apj] {10.1086/158486}, \href
  {http://adsabs.harvard.edu/abs/1980ApJ...242..528E} {242, 528}

\bibitem[\protect\citeauthoryear{{Frerking}, {Langer}  \& {Wilson}}{{Frerking}
  et~al.}{1982}]{Frerking+82}
{Frerking} M.~A.,  {Langer} W.~D.,   {Wilson} R.~W.,  1982, \mn@doi [\apj]
  {10.1086/160451}, \href {http://adsabs.harvard.edu/abs/1982ApJ...262..590F}
  {262, 590}

\bibitem[\protect\citeauthoryear{{Galv{\'a}n-Madrid}
  et~al.,}{{Galv{\'a}n-Madrid} et~al.}{2013}]{Galvan-Madrid+13}
{Galv{\'a}n-Madrid} R.,  et~al., 2013, \mn@doi [\apj]
  {10.1088/0004-637X/779/2/121}, \href
  {http://adsabs.harvard.edu/abs/2013ApJ...779..121G} {779, 121}

\bibitem[\protect\citeauthoryear{{Gibb}, {Jenness}  \& {Economou}}{{Gibb}
  et~al.}{2013}]{Gibb+13}
{Gibb} A.~G.,  {Jenness} T.,   {Economou} F.,  2013, Starlink User Note, \href
  {http://adsabs.harvard.edu/abs/2013StaUN.265.....G} {265}

\bibitem[\protect\citeauthoryear{{Hoare} et~al.,}{{Hoare}
  et~al.}{2012}]{Hoare+12}
{Hoare} M.~G.,  et~al., 2012, \mn@doi [\pasp] {10.1086/668058}, \href
  {http://adsabs.harvard.edu/abs/2012PASP..124..939H} {124, 939}

\bibitem[\protect\citeauthoryear{{Hofner} \& {Churchwell}}{{Hofner} \&
  {Churchwell}}{1996}]{Hofner+Churchwell96}
{Hofner} P.,  {Churchwell} E.,  1996, \aaps, \href
  {http://adsabs.harvard.edu/abs/1996A%26AS..120..283H} {120, 283}

\bibitem[\protect\citeauthoryear{{Hogerheijde}, {van Dishoeck}, {Blake}  \&
  {van Langevelde}}{{Hogerheijde} et~al.}{1998}]{Hogerheijde+98}
{Hogerheijde} M.~R.,  {van Dishoeck} E.~F.,  {Blake} G.~A.,   {van Langevelde}
  H.~J.,  1998, \mn@doi [\apj] {10.1086/305885}, \href
  {http://adsabs.harvard.edu/abs/1998ApJ...502..315H} {502, 315}

\bibitem[\protect\citeauthoryear{{Hou} \& {Han}}{{Hou} \&
  {Han}}{2014}]{Hou+Han14}
{Hou} L.~G.,  {Han} J.~L.,  2014, \mn@doi [\aap] {10.1051/0004-6361/201424039},
  \href {http://adsabs.harvard.edu/abs/2014A%26A...569A.125H} {569, A125}

\bibitem[\protect\citeauthoryear{{Jackson} et~al.,}{{Jackson}
  et~al.}{2006}]{Jackson+06}
{Jackson} J.~M.,  et~al., 2006, \mn@doi [\apjs] {10.1086/500091}, \href
  {http://adsabs.harvard.edu/abs/2006ApJS..163..145J} {163, 145}

\bibitem[\protect\citeauthoryear{{Jenness}, {Cavanagh}, {Economou}  \&
  {Berry}}{{Jenness} et~al.}{2008}]{Jenness+08}
{Jenness} T.,  {Cavanagh} B.,  {Economou} F.,   {Berry} D.~S.,  2008, in
  {Argyle} R.~W.,  {Bunclark} P.~S.,   {Lewis} J.~R.,  eds,  Astronomical
  Society of the Pacific Conference Series Vol. 394, Astronomical Data Analysis
  Software and Systems XVII. p.~565

\bibitem[\protect\citeauthoryear{{Jenness}, {Currie}, {Tilanus}, {Cavanagh},
  {Berry}, {Leech}  \& {Rizzi}}{{Jenness} et~al.}{2015}]{Jenness+15}
{Jenness} T.,  {Currie} M.~J.,  {Tilanus} R.~P.~J.,  {Cavanagh} B.,  {Berry}
  D.~S.,  {Leech} J.,   {Rizzi} L.,  2015, \mn@doi [\mnras]
  {10.1093/mnras/stv1545}, \href
  {http://adsabs.harvard.edu/abs/2015MNRAS.453...73J} {453, 73}

\bibitem[\protect\citeauthoryear{{Johnston}, {Shepherd}, {Aguirre}, {Dunham},
  {Rosolowsky}  \& {Wood}}{{Johnston} et~al.}{2009}]{Johnston+09}
{Johnston} K.~G.,  {Shepherd} D.~S.,  {Aguirre} J.~E.,  {Dunham} M.~K.,
  {Rosolowsky} E.,   {Wood} K.,  2009, \mn@doi [\apj]
  {10.1088/0004-637X/707/1/283}, \href
  {http://adsabs.harvard.edu/abs/2009ApJ...707..283J} {707, 283}

\bibitem[\protect\citeauthoryear{{Kauffmann}, {Bertoldi}, {Bourke}, {Evans}  \&
  {Lee}}{{Kauffmann} et~al.}{2008}]{Kauffmann+08}
{Kauffmann} J.,  {Bertoldi} F.,  {Bourke} T.~L.,  {Evans} II N.~J.,   {Lee}
  C.~W.,  2008, \mn@doi [\aap] {10.1051/0004-6361:200809481}, \href
  {http://adsabs.harvard.edu/abs/2008A%26A...487..993K} {487, 993}

\bibitem[\protect\citeauthoryear{{Kutner} \& {Ulich}}{{Kutner} \&
  {Ulich}}{1981}]{Kutner+Ulich81}
{Kutner} M.~L.,  {Ulich} B.~L.,  1981, \mn@doi [\apj] {10.1086/159380}, \href
  {http://adsabs.harvard.edu/abs/1981ApJ...250..341K} {250, 341}

\bibitem[\protect\citeauthoryear{{Lee}, {Stark}, {Kim}  \& {Moon}}{{Lee}
  et~al.}{2001}]{Lee+01}
{Lee} Y.,  {Stark} A.~A.,  {Kim} H.-G.,   {Moon} D.-S.,  2001, \mn@doi [\apjs]
  {10.1086/321790}, \href {http://adsabs.harvard.edu/abs/2001ApJS..136..137L}
  {136, 137}

\bibitem[\protect\citeauthoryear{{Louvet} et~al.,}{{Louvet}
  et~al.}{2014}]{Louvet+14}
{Louvet} F.,  et~al., 2014, \mn@doi [\aap] {10.1051/0004-6361/201423603}, \href
  {http://adsabs.harvard.edu/abs/2014A%26A...570A..15L} {570, A15}

\bibitem[\protect\citeauthoryear{{Lumsden}, {Hoare}, {Urquhart}, {Oudmaijer},
  {Davies}, {Mottram}, {Cooper}  \& {Moore}}{{Lumsden}
  et~al.}{2013}]{Lumsden+13}
{Lumsden} S.~L.,  {Hoare} M.~G.,  {Urquhart} J.~S.,  {Oudmaijer} R.~D.,
  {Davies} B.,  {Mottram} J.~C.,  {Cooper} H.~D.~B.,   {Moore} T.~J.~T.,  2013,
  \mn@doi [\apjs] {10.1088/0067-0049/208/1/11}, \href
  {http://adsabs.harvard.edu/abs/2013ApJS..208...11L} {208, 11}

\bibitem[\protect\citeauthoryear{{Milam}, {Savage}, {Brewster}, {Ziurys}  \&
  {Wyckoff}}{{Milam} et~al.}{2005}]{Milam+05}
{Milam} S.~N.,  {Savage} C.,  {Brewster} M.~A.,  {Ziurys} L.~M.,   {Wyckoff}
  S.,  2005, \mn@doi [\apj] {10.1086/497123}, \href
  {http://adsabs.harvard.edu/abs/2005ApJ...634.1126M} {634, 1126}

\bibitem[\protect\citeauthoryear{{Molinari} et~al.,}{{Molinari}
  et~al.}{2010a}]{Molinari+10a}
{Molinari} S.,  et~al., 2010a, \mn@doi [\pasp] {10.1086/651314}, \href
  {http://adsabs.harvard.edu/abs/2010PASP..122..314M} {122, 314}

\bibitem[\protect\citeauthoryear{{Molinari} et~al.,}{{Molinari}
  et~al.}{2010b}]{Molinari+10b}
{Molinari} S.,  et~al., 2010b, \mn@doi [\aap] {10.1051/0004-6361/201014659},
  \href {http://adsabs.harvard.edu/abs/2010A%26A...518L.100M} {518, L100}

\bibitem[\protect\citeauthoryear{{Moore}, {Urquhart}, {Morgan}  \&
  {Thompson}}{{Moore} et~al.}{2012}]{Moore+12}
{Moore} T.~J.~T.,  {Urquhart} J.~S.,  {Morgan} L.~K.,   {Thompson} M.~A.,
  2012, \mn@doi [\mnras] {10.1111/j.1365-2966.2012.21740.x}, \href
  {http://adsabs.harvard.edu/abs/2012MNRAS.426..701M} {426, 701}

\bibitem[\protect\citeauthoryear{{Moore} et~al.,}{{Moore}
  et~al.}{2015}]{Moore+15}
{Moore} T.~J.~T.,  et~al., 2015, \mn@doi [\mnras] {10.1093/mnras/stv1833},
  \href {http://adsabs.harvard.edu/abs/2015MNRAS.453.4264M} {453, 4264}

\bibitem[\protect\citeauthoryear{{Motte}, {Schilke}  \& {Lis}}{{Motte}
  et~al.}{2003}]{Motte+03}
{Motte} F.,  {Schilke} P.,   {Lis} D.~C.,  2003, \mn@doi [\apj]
  {10.1086/344538}, \href {http://adsabs.harvard.edu/abs/2003ApJ...582..277M}
  {582, 277}

\bibitem[\protect\citeauthoryear{{Nguyen Luong} et~al.,}{{Nguyen Luong}
  et~al.}{2011}]{Nguyen-Luong+11}
{Nguyen Luong} Q.,  et~al., 2011, \mn@doi [\aap] {10.1051/0004-6361/201016271},
  \href {http://adsabs.harvard.edu/abs/2011A%26A...529A..41N} {529, A41}

\bibitem[\protect\citeauthoryear{{Pettitt}, {Dobbs}, {Acreman}  \&
  {Bate}}{{Pettitt} et~al.}{2015}]{Pettitt+15}
{Pettitt} A.~R.,  {Dobbs} C.~L.,  {Acreman} D.~M.,   {Bate} M.~R.,  2015,
  \mn@doi [\mnras] {10.1093/mnras/stv600}, \href
  {http://adsabs.harvard.edu/abs/2015MNRAS.449.3911P} {449, 3911}

\bibitem[\protect\citeauthoryear{{Polychroni}, {Moore}  \&
  {Allsopp}}{{Polychroni} et~al.}{2012}]{Polychroni+12}
{Polychroni} D.,  {Moore} T.~J.~T.,   {Allsopp} J.,  2012, \mn@doi [\mnras]
  {10.1111/j.1365-2966.2012.20803.x}, \href
  {http://adsabs.harvard.edu/abs/2012MNRAS.422.2992P} {422, 2992}

\bibitem[\protect\citeauthoryear{{Purcell} et~al.,}{{Purcell}
  et~al.}{2013}]{Purcell+13}
{Purcell} C.~R.,  et~al., 2013, \mn@doi [\apjs] {10.1088/0067-0049/205/1/1},
  \href {http://adsabs.harvard.edu/abs/2013ApJS..205....1P} {205, 1}

\bibitem[\protect\citeauthoryear{{Ragan}, {Henning}, {Tackenberg}, {Beuther},
  {Johnston}, {Kainulainen}  \& {Linz}}{{Ragan} et~al.}{2014}]{Ragan+14}
{Ragan} S.~E.,  {Henning} T.,  {Tackenberg} J.,  {Beuther} H.,  {Johnston}
  K.~G.,  {Kainulainen} J.,   {Linz} H.,  2014, \mn@doi [\aap]
  {10.1051/0004-6361/201423401}, \href
  {http://adsabs.harvard.edu/abs/2014A%26A...568A..73R} {568, A73}

\bibitem[\protect\citeauthoryear{{Rosolowsky}, {Pineda}, {Kauffmann}  \&
  {Goodman}}{{Rosolowsky} et~al.}{2008}]{Rosolowsky+08}
{Rosolowsky} E.~W.,  {Pineda} J.~E.,  {Kauffmann} J.,   {Goodman} A.~A.,  2008,
  \mn@doi [\apj] {10.1086/587685}, \href
  {http://adsabs.harvard.edu/abs/2008ApJ...679.1338R} {679, 1338}

\bibitem[\protect\citeauthoryear{{Rosolowsky} et~al.,}{{Rosolowsky}
  et~al.}{2010}]{Rosolowsky+10}
{Rosolowsky} E.,  et~al., 2010, \mn@doi [\apjs] {10.1088/0067-0049/188/1/123},
  \href {http://adsabs.harvard.edu/abs/2010ApJS..188..123R} {188, 123}

\bibitem[\protect\citeauthoryear{{Russeil}}{{Russeil}}{2003}]{Russeil03}
{Russeil} D.,  2003, \mn@doi [\aap] {10.1051/0004-6361:20021504}, \href
  {http://adsabs.harvard.edu/abs/2003A%26A...397..133R} {397, 133}

\bibitem[\protect\citeauthoryear{{Sch{\"o}ier}, {J{\o}rgensen}, {van Dishoeck}
  \& {Blake}}{{Sch{\"o}ier} et~al.}{2002}]{Schoier+02}
{Sch{\"o}ier} F.~L.,  {J{\o}rgensen} J.~K.,  {van Dishoeck} E.~F.,   {Blake}
  G.~A.,  2002, \mn@doi [\aap] {10.1051/0004-6361:20020756}, \href
  {http://adsabs.harvard.edu/abs/2002A%26A...390.1001S} {390, 1001}

\bibitem[\protect\citeauthoryear{{Sch{\"o}ier}, {van der Tak}, {van Dishoeck}
  \& {Black}}{{Sch{\"o}ier} et~al.}{2005}]{Schoier+05}
{Sch{\"o}ier} F.~L.,  {van der Tak} F.~F.~S.,  {van Dishoeck} E.~F.,   {Black}
  J.~H.,  2005, \mn@doi [\aap] {10.1051/0004-6361:20041729}, \href
  {http://esoads.eso.org/abs/2005A%26A...432..369S} {432, 369}

\bibitem[\protect\citeauthoryear{{Schuller} et~al.,}{{Schuller}
  et~al.}{2009}]{Schuller+09}
{Schuller} F.,  et~al., 2009, \mn@doi [\aap] {10.1051/0004-6361/200811568},
  \href {http://adsabs.harvard.edu/abs/2009A%26A...504..415S} {504, 415}

\bibitem[\protect\citeauthoryear{{Scoville}, {Sargent}, {Sanders}, {Claussen},
  {Masson}, {Lo}  \& {Phillips}}{{Scoville} et~al.}{1986}]{Scoville+86}
{Scoville} N.~Z.,  {Sargent} A.~I.,  {Sanders} D.~B.,  {Claussen} M.~J.,
  {Masson} C.~R.,  {Lo} K.~Y.,   {Phillips} T.~G.,  1986, \mn@doi [\apj]
  {10.1086/164086}, \href {http://adsabs.harvard.edu/abs/1986ApJ...303..416S}
  {303, 416}

\bibitem[\protect\citeauthoryear{{Sievers}, {Mezger}, {Bordeon}, {Kreysa},
  {Haslam}  \& {Lemke}}{{Sievers} et~al.}{1991}]{Sievers+91}
{Sievers} A.~W.,  {Mezger} P.~G.,  {Bordeon} M.~A.,  {Kreysa} E.,  {Haslam}
  C.~G.~T.,   {Lemke} R.,  1991, \aap, \href
  {http://adsabs.harvard.edu/abs/1991A%26A...251..231S} {251, 231}

\bibitem[\protect\citeauthoryear{{Smith}, {Whitney}, {Conti}, {de Pree}  \&
  {Jackson}}{{Smith} et~al.}{2009}]{Smith+09}
{Smith} N.,  {Whitney} B.~A.,  {Conti} P.~S.,  {de Pree} C.~G.,   {Jackson}
  J.~M.,  2009, \mn@doi [\mnras] {10.1111/j.1365-2966.2009.15343.x}, \href
  {http://adsabs.harvard.edu/abs/2009MNRAS.399..952S} {399, 952}

\bibitem[\protect\citeauthoryear{{Stark} \& {Lee}}{{Stark} \&
  {Lee}}{2006}]{Stark+Lee06}
{Stark} A.~A.,  {Lee} Y.,  2006, \mn@doi [\apjl] {10.1086/504036}, \href
  {http://adsabs.harvard.edu/abs/2006ApJ...641L.113S} {641, L113}

\bibitem[\protect\citeauthoryear{{Steiman-Cameron}, {Wolfire}  \&
  {Hollenbach}}{{Steiman-Cameron} et~al.}{2010}]{Steiman-Cameron+10}
{Steiman-Cameron} T.~Y.,  {Wolfire} M.,   {Hollenbach} D.,  2010, \mn@doi
  [\apj] {10.1088/0004-637X/722/2/1460}, \href
  {http://adsabs.harvard.edu/abs/2010ApJ...722.1460S} {722, 1460}

\bibitem[\protect\citeauthoryear{{Stenholm}}{{Stenholm}}{1975}]{Stenholm75}
{Stenholm} B.,  1975, \aap, \href
  {http://adsabs.harvard.edu/abs/1975A%26A....39..307S} {39, 307}

\bibitem[\protect\citeauthoryear{{Stutzki}, {Bensch}, {Heithausen}, {Ossenkopf}
   \& {Zielinsky}}{{Stutzki} et~al.}{1998}]{Stutzki+98}
{Stutzki} J.,  {Bensch} F.,  {Heithausen} A.,  {Ossenkopf} V.,   {Zielinsky}
  M.,  1998, \aap, \href {http://adsabs.harvard.edu/abs/1998A%26A...336..697S}
  {336, 697}

\bibitem[\protect\citeauthoryear{{Taylor} \& {Cordes}}{{Taylor} \&
  {Cordes}}{1993}]{Taylor+Cordes93}
{Taylor} J.~H.,  {Cordes} J.~M.,  1993, \mn@doi [\apj] {10.1086/172870}, \href
  {http://adsabs.harvard.edu/abs/1993ApJ...411..674T} {411, 674}

\bibitem[\protect\citeauthoryear{{Urquhart} et~al.,}{{Urquhart}
  et~al.}{2013}]{Urquhart+13b}
{Urquhart} J.~S.,  et~al., 2013, \mn@doi [\mnras] {10.1093/mnras/stt1310},
  \href {http://adsabs.harvard.edu/abs/2013MNRAS.435..400U} {435, 400}

\bibitem[\protect\citeauthoryear{{Urquhart}, {Figura}, {Moore}, {Hoare},
  {Lumsden}, {Mottram}, {Thompson}  \& {Oudmaijer}}{{Urquhart}
  et~al.}{2014}]{Urquhart+14}
{Urquhart} J.~S.,  {Figura} C.~C.,  {Moore} T.~J.~T.,  {Hoare} M.~G.,
  {Lumsden} S.~L.,  {Mottram} J.~C.,  {Thompson} M.~A.,   {Oudmaijer} R.~D.,
  2014, \mn@doi [\mnras] {10.1093/mnras/stt2006}, \href
  {http://adsabs.harvard.edu/abs/2014MNRAS.437.1791U} {437, 1791}

\bibitem[\protect\citeauthoryear{{Williams}, {de Geus}  \& {Blitz}}{{Williams}
  et~al.}{1994}]{Williams+94}
{Williams} J.~P.,  {de Geus} E.~J.,   {Blitz} L.,  1994, \mn@doi [\apj]
  {10.1086/174279}, \href {http://adsabs.harvard.edu/abs/1994ApJ...428..693W}
  {428, 693}

\bibitem[\protect\citeauthoryear{{Wilson} \& {Rood}}{{Wilson} \&
  {Rood}}{1994}]{Wilson+Rood94}
{Wilson} T.~L.,  {Rood} R.,  1994, \mn@doi [\araa]
  {10.1146/annurev.aa.32.090194.001203}, \href
  {http://adsabs.harvard.edu/abs/1994ARA%26A..32..191W} {32, 191}

\bibitem[\protect\citeauthoryear{{Wright} et~al.,}{{Wright}
  et~al.}{2010}]{Wright+10}
{Wright} E.~L.,  et~al., 2010, \mn@doi [\aj] {10.1088/0004-6256/140/6/1868},
  \href {http://adsabs.harvard.edu/abs/2010AJ....140.1868W} {140, 1868}

\bibitem[\protect\citeauthoryear{{Zhang}, {Reid}, {Menten}, {Zheng},
  {Brunthaler}, {Dame}  \& {Xu}}{{Zhang} et~al.}{2013}]{Zhang+13}
{Zhang} B.,  {Reid} M.~J.,  {Menten} K.~M.,  {Zheng} X.~W.,  {Brunthaler} A.,
  {Dame} T.~M.,   {Xu} Y.,  2013, \mn@doi [\apj] {10.1088/0004-637X/775/1/79},
  \href {http://cdsads.u-strasbg.fr/abs/2013ApJ...775...79Z} {775, 79}

\bibitem[\protect\citeauthoryear{{Zhang} et~al.,}{{Zhang}
  et~al.}{2014}]{Zhang+14}
{Zhang} B.,  et~al., 2014, \mn@doi [\apj] {10.1088/0004-637X/781/2/89}, \href
  {http://adsabs.harvard.edu/abs/2014ApJ...781...89Z} {781, 89}

\bibitem[\protect\citeauthoryear{{de Vaucouleurs} \& {Malik}}{{de Vaucouleurs}
  \& {Malik}}{1969}]{deVaucouleurs+Malik69}
{de Vaucouleurs} G.,  {Malik} G.~M.,  1969, \mnras, \href
  {http://adsabs.harvard.edu/abs/1969MNRAS.142..387D} {142, 387}

\makeatother
\end{thebibliography}


\newpage
\appendix

\section{{\sc orac-dr} parameters}\label{app:orac-pars}

The following parameters were used in the reduction of each CHIMPS observation:\\

\noindent \texttt{
{[}REDUCE\_SCIENCE\_NARROWLINE{]}\\
BASELINE\_LINEARITY=1\\
BASELINE\_ORDER=4\\
BASELINE\_LINEARITY\_LINEWIDTH=$^{1}$\\
PIXEL\_SCALE=7.612\\
REBIN=0.5\\
}

$^{1}$This parameter was altered for each individual observation, and was determined by inspecting the average spectrum of each raw time series file. For example, if a large observation feature was visible with a 10\kms\ linewidth at a velocity of 40\kms, then this parameter would be set to as  `{\texttt 30:50}'.

\section{FellWalker parameters}\label{app:fwpars}

The source extraction was carried out using the FellWalker algorithm \citep{Berry15} of  {\sc cupid:findclumps}, and the SNR cubes were the input data. Pixels with an SNR $<$ 3 are regarded as noise, and all clumps must have a peak SNR of 8 on top of the noise level. The FellWalker parameters are listed below. The \coa\ source extraction parameters were:

\noindent\texttt{\\
FellWalker.AllowEdge=0 \\
FellWalker.CleanIter=2\\
FellWalker.FlatSlope=1\\
FellWalker.FwhmBeam=3\\
FellWalker.MaxBad=0.05\\
FellWalker.MinDip=5\\
FellWalker.MinHeight=5\\
FellWalker.MinPix=16\\
FellWalker.MaxJump=4\\
FellWalker.Noise=3\\
FellWalker.RMS=1\\
FellWalker.VeloRes=1.}
\\

\noindent An identical parameter file was used for the \cob\ source extraction with the following alterations:

\noindent\texttt{\\
FellWalker.MinDip=3\\
FellWalker.MinHeight=5\\
FellWalker.Noise=2}.

\bsp	
\label{lastpage}
\end{document}